\documentclass[default,iicol]{sn-jnl}% Default with double column layout

\usepackage{graphicx}%
\usepackage{multirow}%
\usepackage{amsmath,amssymb,amsfonts}%
\usepackage{amsthm}%
\usepackage{mathrsfs}%
\usepackage[title]{appendix}%
\usepackage{xcolor}%
\usepackage{textcomp}%
\usepackage{manyfoot}%
\usepackage{booktabs}%
\usepackage{algorithm}%
\usepackage{algorithmicx}%
\usepackage{algpseudocode}%
\usepackage{listings}%
\usepackage{comment}%
\usepackage{hyperref} 
\theoremstyle{thmstyleone}%
\theoremstyle{thmstyletwo}%
\theoremstyle{thmstylethree}%
\begin{document}

\title[Machine Learning for the Production of Official Statistics: Density Ratio Estimation using Biased Transaction Data for Japanese labor statistics]{Machine Learning for the Production of Official Statistics: Density Ratio Estimation using Biased Transaction Data for Japanese labor statistics
}

\author*[1,2]{\fnm{Yuya} \sur{Takada}}\email{yuyatakada@g.ecc.u-tokyo.ac.jp}
\author[1]{\fnm{Kiyoshi} \sur{Izumi}}\email{izumi@sys.t.u-tokyo.ac.jp}

\affil[1]{\orgdiv{Department of Systems Innovation, School of Engineering}, \orgname{The University of Tokyo}, \orgaddress{\state{Tokyo}, \country{Japan}}}
\affil[2]{\orgname{Re Data Science Co., Ltd.}, \orgaddress{\state{Chiba}, \country{Japan}}}

\abstract{
National statistical institutes are beginning to use non-traditional data sources to produce official statistics. These sources, originally collected for non-statistical purposes, include point-of-sales(POS) data and mobile phone global positioning system(GPS) data. Such data have the potential to significantly enhance the usefulness of official statistics. In the era of big data, many private companies are accumulating vast amounts of transaction data. Exploring how to leverage these data for official statistics is increasingly important. However, progress has been slower than expected, mainly because such data are not collected through sample-based survey methods and therefore exhibit substantial selection bias. If this bias can be properly addressed, these data could become a valuable resource for official statistics, substantially expanding their scope and improving the quality of decision-making, including economic policy. This paper demonstrates that even biased transaction data can be useful for producing official statistics for prompt release, by drawing on the concepts of density ratio estimation and supervised learning under covariate shift, both developed in the field of machine learning. As a case study, we show that preliminary statistics can be produced in a timely manner using biased data from a Japanese private employment agency. This approach enables the early release of a key labor market indicator that would otherwise be delayed by up to a year, thereby making it unavailable for timely decision-making.
}

\maketitle

\section{Introduction}\label{sec1}
In recent years, national statistical institutes have started using non-traditional data sources to produce official statistics.
These new data sources are not directly related to statistical production. Such data are collected for purposes other than official statistics. For example, point-of-sales(POS) data registered in retail stores, real-time population data captured by the global positioning system(GPS) of mobile phones, price data collected from hundreds of online retailers, satellite imagery data, and internet search data such as those published in Google Trends can also be used to create official statistics. 

These new data sources have the potential to substantially enhance the utility of official statistics. If economic conditions can be captured more comprehensively and in a timelier manner by the official statistics made by these new data sources, it is reasonable to expect that the quality of decision-making, including economic policy, will improve.

In the era of big data, many private companies accumulate vast amounts of transaction data. It is imperative to explore ways to utilize these data sources for official statistics. However, progress in utilizing such data has been slower than anticipated. This is primarily because most of these data cannot be directly employed for statistical purposes. Since they are not collected through survey methodologies based on sample design, they exhibit substantial selection bias when considered as a source for official statistics. In other words, if this strong selection bias can be appropriately corrected, a substantial volume of potentially valuable transaction data could be utilized as a source for official statistics. This would significantly expand the scope of official statistics and enhance the quality of various decision-making processes.

This study aims to develop a framework for producing official statistics using the new data sources that are subject to selection bias, particularly transaction data collected for purposes other than official statistics and not obtained through survey-based methodologies. Specifically, we propose a framework for producing preliminary reports based on partial data that, while prone to selection bias, offers high timeliness. This approach is designed for indicators where comprehensive nationwide data has already been collected for official statistics, yet the substantial time lag before publication limits its utility for timely decision-making. In other words, this study focuses on the speed of publication as one of the advantages of using new data sources. Surveys designed to produce official statistics are time-consuming because survey makers need to collect survey responses from firms or households. On the other hand, we can obtain such new data sources with almost no time lag as the previous day’s information is available the next day. If the proposed framework enables official statistics to capture economic conditions more promptly, it is reasonable to anticipate improvements in the quality of decision-making, including economic policy formulation.

The approach of utilizing these components with selection bias to infer the characteristics of the whole is a well-recognized challenge in the field of machine learning, and extensive research has been conducted on this topic. To put it another way, integrating insights from machine learning into the production of official statistics has the potential to fundamentally address the current challenges faced by official statistics. In this study, we focus on methodologies developed in the field of machine learning, specifically Density ratio estimation and the concept of covariate shift.

As an experiment, we showed that it is possible to publish prompt preliminary statistics using biased data from a Japanese private employment agency.

Policymakers, business managers, and recruitment professionals must know labor market trends as soon as possible for their decision-making. For this purpose, a range of official statistics is provided in many countries, offering insights into unemployment rates, workforce numbers, average wages, and so on, typically with a lag of only two to three months. 
However, some indicators are difficult to grasp promptly and have a lag of more than a year. A prime example is the pressure on wages due to supply and demand in the external labor market. The prompt publication of the indicator would be useful for decision-making by policymakers and those involved in corporate management and recruitment.

Although there are several possible patterns of indicators for capturing the pressure on wages due to supply and demand in the external labor market, in this case, we adopt the wage changes of career-changing employees: the proportion of individuals with an increased wage after changing careers. Within the given period, the total number of job switchers serves as the denominator, and the subset of these individuals who experienced a wage increase exceeding $10\%$ post-career change constitutes the numerator. The expression over $10\%$ increased refers to a marked increase; consequently, the indicator represents the percentage of individuals whose wages increased by more than 10 percent after they switched careers. In official Japanese statistics, the indicator is published with a time lag of 6 to 13 months. The challenge this time is to eliminate this time lag, which could be one year or longer.

Job changes in the external labor market occur through various channels: job advertisements, public and private employment agencies, and employee referrals. In this study, we focused on transaction data held by private employment agencies because they hold such data in a form that can be used in real-time. Private employment agencies interview job seekers for accurate wage information before they change jobs. In addition, during the process of a job seeker deciding to change jobs, the employment agency calculates and shows the job seeker the annual wages she or he will earn if she or he changes jobs. In other words, they inevitably possess data on wages both before and after a job change. However, in the process of changing jobs via job advertisements or referrals, there is no process of recording wages before or after the job change in some database. In addition, public employment agencies can in principle capture pre- and post-job change wage data and use it for statistical purposes; however, in the case of Japan, this system is not fully implemented at present.

We obtained transaction data from Japan's largest private employment agency. In addition, the Japanese government provided us with anonymized sample data on the official statistics to be benchmarked, which are not available to the general public. Both are raw data before tabulation, rather than so called statistical data after tabulation. Much of the processing in this study is done at the level of individual raw data, rather than post-aggregated statistical data. It should be emphasized that individuals cannot be identified and that issues in terms of data protection and other aspects are dealt with very carefully.

Transaction data held by private employment agencies is available in real time, but the coverage rate is quite low - less than $5\%$ even for the largest agency case used in this study - and furthermore, when considered as a sample to understand the whole of Japan, it has a very strong bias. As an illustrative example, in the second half of 2018, the average age of the population in the official statistics—namely, full-time employees who changed jobs—was approximately 40 years. In contrast, the corresponding figure in the transaction data was around 31 years. Similarly, the proportion of individuals with a university degree or higher was about $35\%$ in the official statistics, whereas it was approximately $79\%$ in the transaction data, indicating a substantial discrepancy between the two datasets.

This is because the sample is limited to those who have changed jobs through the private employment agency. In other words, the sample is limited to those whom companies hiring them would like to hire even if they have to pay a fee to the private employment agency. On the other hand, official statistics, as mentioned above, have a long time lag before publication, but capture Japan as a whole. In other words, if we can learn the relationship between transaction data held by private employment agencies and official statistics, and use this information to remove bias on the private side, we can solve this problem. In this study, we show this using an application of the density ratio estimation and covariate shift concept.

The contribution of this study is not limited to the labor statistics used in the experiment. The approach introduced in this research is adaptable to not just these labor statistics but also to a range of other data sets. There are thought to be many situations with the same structure.

\section{Related Work}\label{sec2}
As mentioned in the previous section, recently, the national statistical institutes have started using non-traditional data sources to produce official statistics. The new data sources do not have a direct connection to the aims of statistical production. The transaction data of the private employment agencies used in this study fall into this category, but there are some other examples. Previous studies have documented cases in which a variety of data — including satellite imagery, GPS-based population data from mobile phones, POS records from retail outlets, online price data, and internet search trends such as Google Trends — have been utilized in the production of official statistics.

The ITU, a United Nations entity focused on information and communication technology, is pioneering the utilization of GPS-based real-time population data via mobile phones. As the Mobile Data Task Team's secretariat within the United Nations statistics division, ITU enhances the use of such data globally \cite{bib8}. Between 2016 and 2018, in countries such as Colombia, Georgia, Kenya, the Philippines, Sweden, and the UAE, local government bodies gathered and processed telecom provider data to augment household surveys and official records, resulting in 16 different indicators. During 2020-2021, Brazil and Indonesia's investigations leveraged telecom data to analyze two Sustainable Development Goal indicators: mobile network coverage and internet access. Furthermore, the ITU released a guide entitled Handbook on the Use of Mobile Phone Data for Official Statistics \cite{bib9} to facilitate the application of mobile data. In Estonia, budget limitations led to the replacement of traditional travel surveys with mobile SIM card data for tracking travel expenditures \cite{bib11}.

Since the early 2000s, the application of POS data has expanded significantly across multiple countries. Notably, in Switzerland, Norway, and the Netherlands, there has been a focused effort to utilize this data to mitigate issues associated with conventional data sources. Moreover, beginning with the study referenced as \cite{bib1}, novel price indices were developed using POS data, with their attributes being confirmed in the studies \cite{bib2} and \cite{bib3}. 
In Japan, a corporate entity is engaged in the provision of price indices services, which are created from POS data following the methodology described in \cite{bib4}. The Bank of Japan incorporates these indices in its various reports. Furthermore, price indices have been established using data on prices gathered from a multitude of online retailers. The Billion Prices Project, an academic initiative, commenced in 2008 by Professors Alberto Cavallo and Roberto Rigobon at MIT Sloan and Harvard Business School. In this context, new price indices, derived from data collected from retailers \cite{bib5, bib6}, were introduced, alongside a novel approach for developing purchasing power parities (PPPs) \cite{bib7}.

The Organisation for Economic Co-operation and Development (OECD) utilizes internet search data from Google Trends. This data aids in the publication of the OECD Weekly Tracker of Economic Activity, which provides insights into the weekly Gross Domestic Product (GDP) of 46 countries, including those in the OECD and G20 \cite{bib13}. Additionally, Google Trends data were used in estimating unemployment rates \cite{bib14}. Google Trends data is regarded as an auxiliary series for the estimation of monthly unemployment figures with the official labor force survey.

Recent studies in the Journal of Computational Social Science have utilized non-traditional data sources to address the limitations of traditional economic indicators and offer complementary insights into socio-economic dynamics. For instance, \cite{bib38} examined the relationship between media sentiment in press articles and traditional economic indicators—PMI, CCI, and employment—during the COVID-19 period in Poland. Their findings suggest that media sentiment can serve as a leading indicator for these economic metrics, with varying lead times. Similarly, \cite{bib39} investigated the use of publicly available organic data, such as Google Trends and Twitter, to predict forced migration from Ukraine during the 2022 refugee crisis. They found that certain digital indicators could effectively forecast migration flows into neighboring countries. \cite{bib40} combined Twitter and mobile phone data to observe cross-border mobility during the Turkish-European border opening in 2020. Their study highlighted the benefits and limitations of these data sources in capturing real-time migration patterns. Lastly, \cite{bib40} analyzed international mobility between the UK and Europe around Brexit by integrating official statistics with non-traditional data sources, including scientific publications and air passenger data. Their comprehensive approach provided nuanced insights into migration trends influenced by geopolitical changes.

In this study, we demonstrated that even data with selection bias can be valuable for producing official statistics by applying density ratio estimation and the concept of covariate shift. As an experiment, we showed that it is possible to publish prompt preliminary statistics for labor market using biased data from a private employment agency. A previous study used a similar approach in \cite{bib33}. This previous study assumes that government agencies can obtain data from private employment agencies in real time and combine them with raw data from the most recent official statistics to complete the estimation work within the government. However, this study does not make this unrealistic assumption. In the case of Japan, it is challenging for that government agencies to obtain data from private employment agencies in real time and finish the estimation work within the government. Realistically, a separate organization outside the government agency would need to combine the real-time private-sector employment agency data with the raw data from the most recent official statistics available and complete the estimation work. In that case, the time lag in the raw data of available government statistics would be greatly increased. In this study, we show that estimating with sufficient accuracy is possible in this situation.

\section{Task Setting}\label{sec3}
In this section, we describe the structure of the task, the evaluation method, and clarify how the experiment conducted in this study is positioned within the typology of selection bias. To maintain a certain level of generality in the discussion, we do not delve into the detailed experimental settings or data description here; these specific details will be provided in Section \ref{sec5}, following the explanation of the methodology in Section \ref{sec4}.

\subsection{Structure of the Task}\label{sec31}
\begin{figure*}[htbp]
\centering
\includegraphics[width=1.0\textwidth]{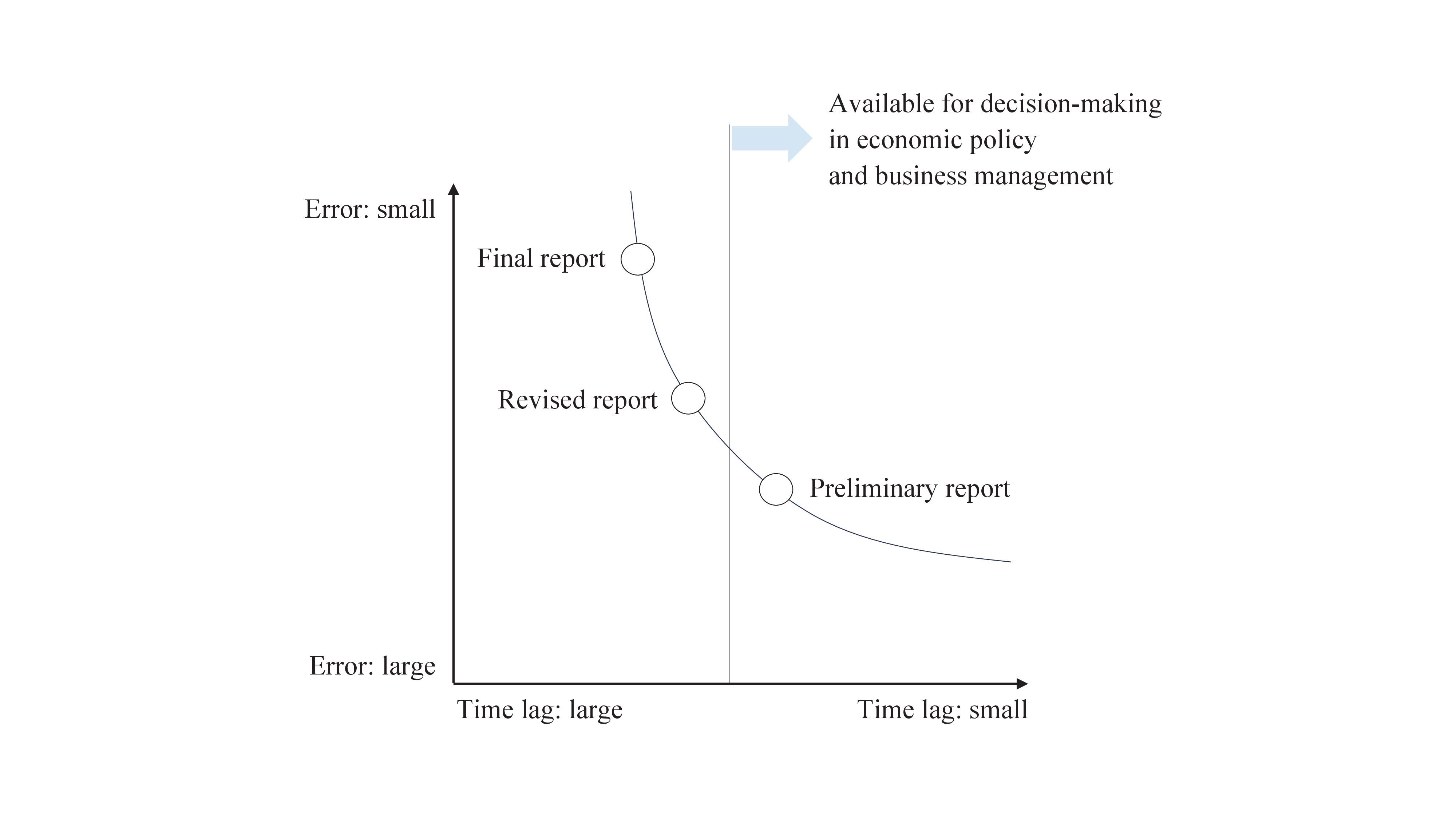}
\caption{Errors and time lags in official statistics}
\label{Fig1}
\end{figure*}

Typical official statistics are released multiple times in stages, such as preliminary, revised, and final reports. Naturally, the accuracy of earlier releases tends to be lower, while later releases are more precise. 
In most cases, only the preliminary official statistical indicators are available for use in economic policy and business decision-making. This structure is illustrated in Fig. \ref{Fig1}. In Fig. \ref{Fig1}, the term error schematically represents the magnitude of the discrepancy between the true statistical value—which is fundamentally unobservable—and the value actually captured through the survey.

\begin{figure*}[htbp]
\centering
\includegraphics[width=1.0\textwidth]{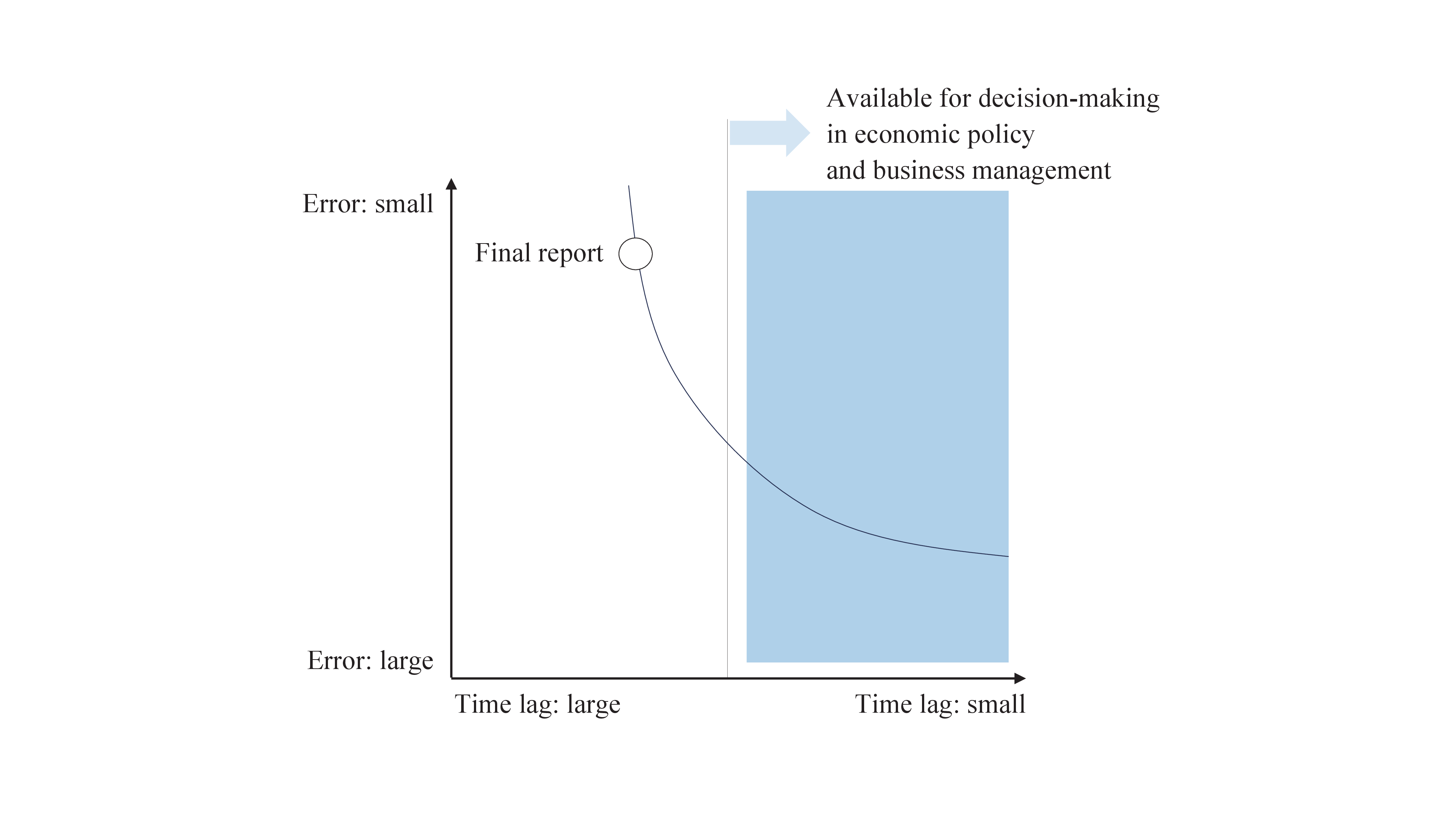}
\caption{Targeted scope of this study}
\label{Fig2}
\end{figure*}

However, many official statistical indicators do not have preliminary releases that can be used for decision-making, even when these indicators are of great importance. This is often due to challenges such as the high cost of surveys or the burden imposed on private enterprises, making rapid data collection and publication difficult. This study focuses on such indicators. As illustrated in Fig. \ref{Fig2}, this can be understood as an effort to develop the shadowed section on the right side.

\subsection{Evaluation}\label{sec32}
In Fig. \ref{Fig1} and \ref{Fig2}, the term error schematically represents the magnitude of the discrepancy between the true statistical value—which is fundamentally unobservable—and the value actually captured through the survey. However, in this study, it is not feasible to evaluate the preliminary estimates along this axis, as the true statistical values are unobservable. Therefore, we treat the final estimates as the ground truth and assess the quality of the preliminary estimates based on the magnitude of their difference between the estimated and final values. To measure the performance of our estimation, we used the mean absolute error (MAE). In this experiment, therefore, MAE is calculated by comparing our estimated values of the proportion of individuals with increased wages after changing careers to the values published by the government.

For MAE, a smaller value is generally preferable; however, there is no established standard stipulating that meeting this criterion is sufficient for an indicator to be considered desirable as a prompt preliminary official indicator. Even official prompt preliminary indicators released by government agencies often have errors that cannot be considered small. This does not present a fundamental issue, provided that users are aware of these errors and utilize the indicators with appropriate caution. Despite the presence of errors that cannot be considered small, such indicators can enhance the quality of decision-making compared to scenarios where they are unavailable altogether.

However, it goes without saying that indicators with large errors caused by substantial selection bias can only be used in a limited range of practical situations. As discussed in Section \ref{sec1}, this is one of the reasons why the use of transaction data held by private companies has not advanced as much as expected.

\begin{figure*}[htbp]
\centering
\includegraphics[width=1.0\textwidth]{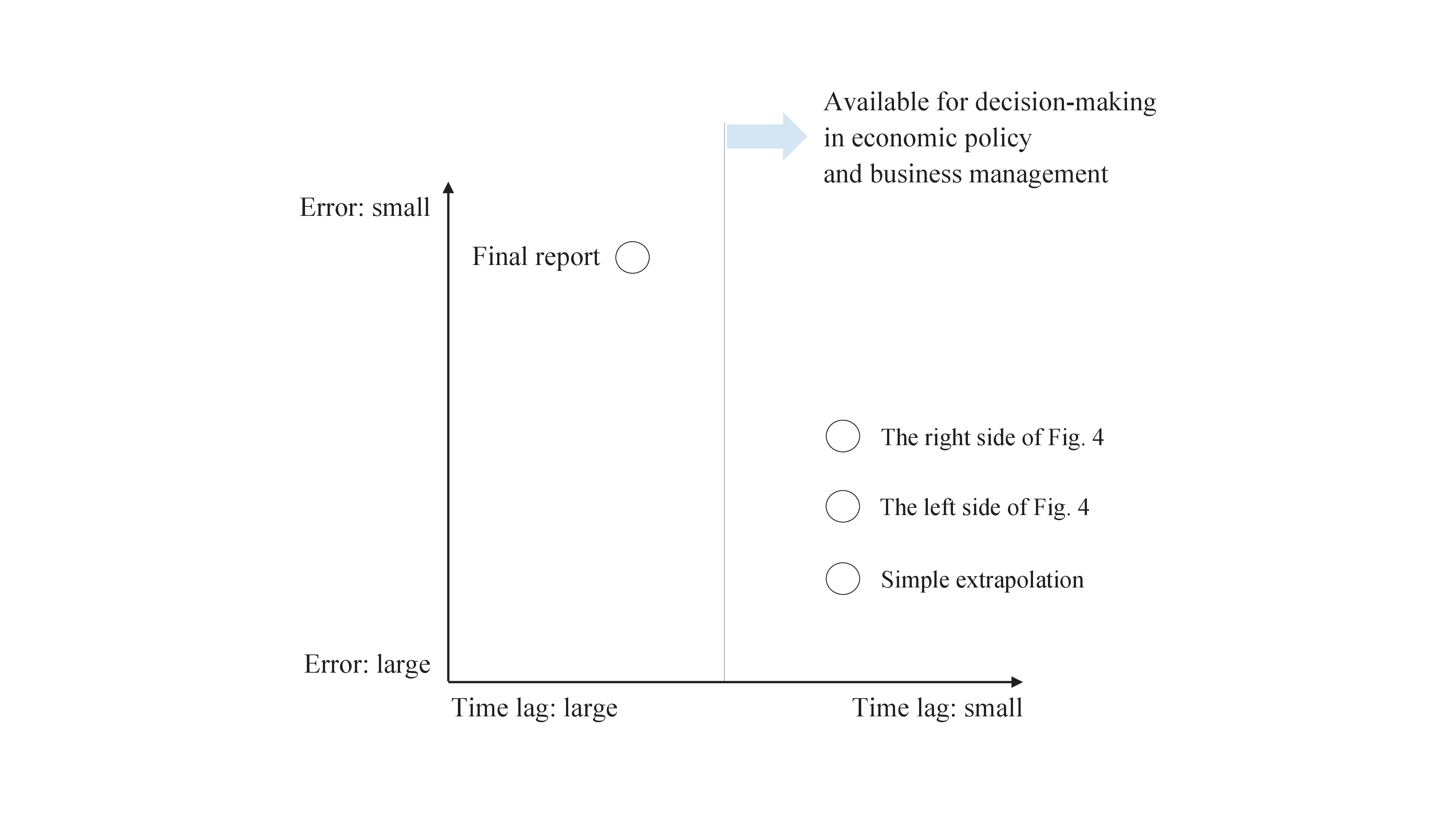}
\caption{Structure of the Experimental Design}
\label{Fig3}
\end{figure*}

In this study, building on this, we aim to evaluate the extent to which accuracy is improved through the mitigation of selection bias. Therefore, the simple extrapolation-based prediction, which, although informative, does not incorporate any correction for selection bias, is treated as a benchmark, and we first examine whether the estimates proposed in Sections \ref{sec42}, \ref{sec43}, and \ref{sec46}, as shown on the left side of Fig. \ref{Fig4} in Section \ref{sec41}, can surpass it. Subsequently, we demonstrate that the estimation methods integrating the additional process proposed in Sections \ref{sec44} and \ref{sec45}, as shown on the right side of Fig. \ref{Fig4} in Section \ref{sec41}, improve in accuracy compared to those without the additional process. This structure is illustrated in Fig. \ref{Fig3}. As in Fig. \ref{Fig1} and Fig. \ref{Fig2}, Fig. \ref{Fig3} uses the term error to schematically represent the magnitude of the discrepancy between the true statistical value—which is fundamentally unobservable—and the value actually captured through the survey. However, in this study, it should be noted that the evaluation is conducted by assessing the quality of the preliminary estimates based on the magnitude of their difference from the final report published by the government, since the true statistical values themselves are not observable.

Specifically, the following simple extrapolation-based prediction was used as a benchmark.

\begin{equation}\label{Simple}
\widehat{o_{T}} = o_{T-2} \dfrac{s_{T}}{s_{T-2}},
\end{equation}

where $\widehat{o_{T}}$ is the official statistics value for the period $T$ , which is the target of extrapolation-based prediction, and $o_{T-2}$ is the actual value of the official statistics for the $T-2$ period. $s_{T}$ and $s_{T-2}$ are the supplementary indicators for the $T$ and $T-2$ periods, respectively. In the empirical setting considered in this study, both $o$ and $s$ are indicators showing the percentage of individuals who increased their wages by $10\%$ or more upon changing jobs.

In relation to the notation used in the subsequent Section \ref{sec4}, $o_{T}$ is equivalent to the total number of components that indicate $1$ in the vector $Y_{g,T}$, divided by the number of dimensions. Similarly, $s_{T}$ is equivalent to the total number of components that indicate $1$ in the vector $Y_{p,T}$ when the classification label is adopted, divided by the number of dimensions.

We conducted the Harvey, Leybourne, and Newbold (HLN) test to evaluate whether the performance of the methods proposed in this study exceeded that of the simple extrapolation method. The power parameter utilized in the loss function was set to 1, corresponding to MAE.

\subsection{Types of Selection Bias and Position of this Study}\label{sec33}
The task we address can be regarded as arising under a covariate shift setting. Covariate shift is a situation wherein training and test input points follow different probability distributions but the conditional distributions of output values of given input points are unchanged, which is often encountered in machine learning \cite{bib17}.

Let $\mathcal{D} \subset \mathbb{R}^{d}$ be the domain of covariates, where $d$ is a positive integer.
Suppose $x \in \mathcal{D}$ and $y \in \mathcal{D}' \subset \mathbb{R}$ denote a covariate and its class label, respectively. $p_{S}$ and $p_{T}$ denotes the training and test probability distribution, respectively.
Under these definitions, the setting is as follows:

\begin{equation}\label{covariate shift1} 
p_{S}(y \mid x) = p_{T}(y \mid x)     \: and \:  p_{S}(x) \neq p_{T}(x).
\end{equation}

Although standard learning methods such as maximum likelihood estimation are biased under covariate shift, we can correct the bias asymptotically by weighting the loss function according to the density ratio \cite{bib28}.

The situation referred to as covariate shift can be regarded as a particular class of selection bias. Since \cite{bib29}, it has become customary to handle sample selection bias by dividing missing data mechanisms into three categories: MCAR, MAR, and NMAR.

\begin{itemize}
\item MCAR: Missing completely at random refers to a situation where the probability of data being missing is unrelated to the specific value that should be obtained or the set of observed responses.

\item MAR: Missing at random is a more realistic condition where the probability of responses being missing depends on the set of observed responses but not on the specific missing values.

\item NMAR: Not missing at random is a more challenging situation where the missing data pattern is non-random and depends on the missing variables.
\end{itemize}

According to \cite{bib30}, the covariate shift assumption is considered equivalent to the MAR assumption. However, as \cite{bib31} points out, MAR and NMAR are not fundamentally distinct but rather exist on a continuum, and in this study, we consider the situation we address to be more appropriately regarded as NMAR.

In the NMAR case, \cite{bib32} states that the response probability cannot be verified using only the observed study variables, and therefore additional assumptions are often required. The correction method outlined in Section \ref{sec462} corresponds to this additional assumption.

\section{Methodology}\label{sec4}
In this section, we first describe the architecture of the estimation in Section \ref{sec41}. Then, we present the detailed estimation process from Section \ref{sec42} to Section \ref{sec46}.

\subsection{Architecture of the Estimation}\label{sec41}
\begin{figure*}[htbp]
\centering
\includegraphics[width=0.9\textwidth]{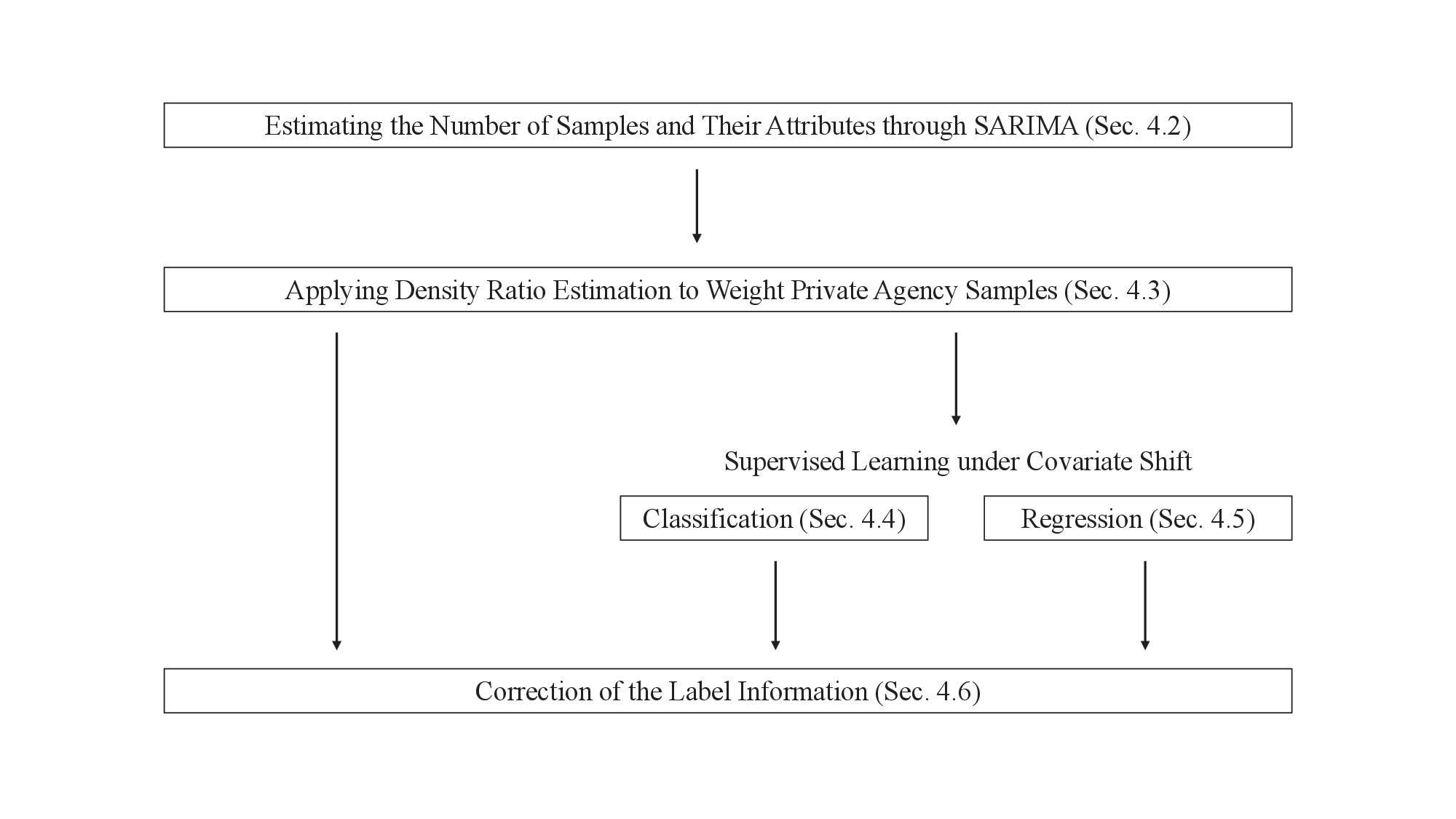}
\caption{Architecture of the Estimation}
\label{Fig4}
\end{figure*}

Fig. \ref{Fig4} illustrates the architecture of the estimation proposed in this paper. Initially, in Section \ref{sec42}, we describe the process of estimating the number of samples and their attributes through SARIMA. Here, we used only historical hired career-changing employee samples from the government survey to estimate the number of samples for the target half-year by career-changing channels, and obtained attribute information for each sample using SARIMA. Second, in Section \ref{sec43}, we explain how to apply density ratio estimation to weight private employment agency samples. Density ratio estimation was performed using both private employment agency transaction data and historical samples estimated from the government survey in the previous step. Third, we describe the step of supervised learning under covariate shift with classification in Section \ref{sec44} and regression in Section \ref{sec45}. Supervised learning under covariate shift was conducted using weighted private employment agency samples obtained in the second step. We could skip the third step since samples with label information for the target half-year were already obtained in the second step, as shown on the left side of Fig. \ref{Fig4}. The purpose of the third step was to reduce errors. In Section \ref{sec46}, we explain the step of correction of the label information. We corrected the label information of each sample estimated in the second/third step and calculated wage changes of hired career-changing employees: the proportion of individuals whose wages increased after changing careers.

\subsection{Estimating the Number of Samples and Their Attributes through SARIMA}\label{sec42}
\begin{figure*}[htbp]
\centering
\includegraphics[width=0.9\textwidth]{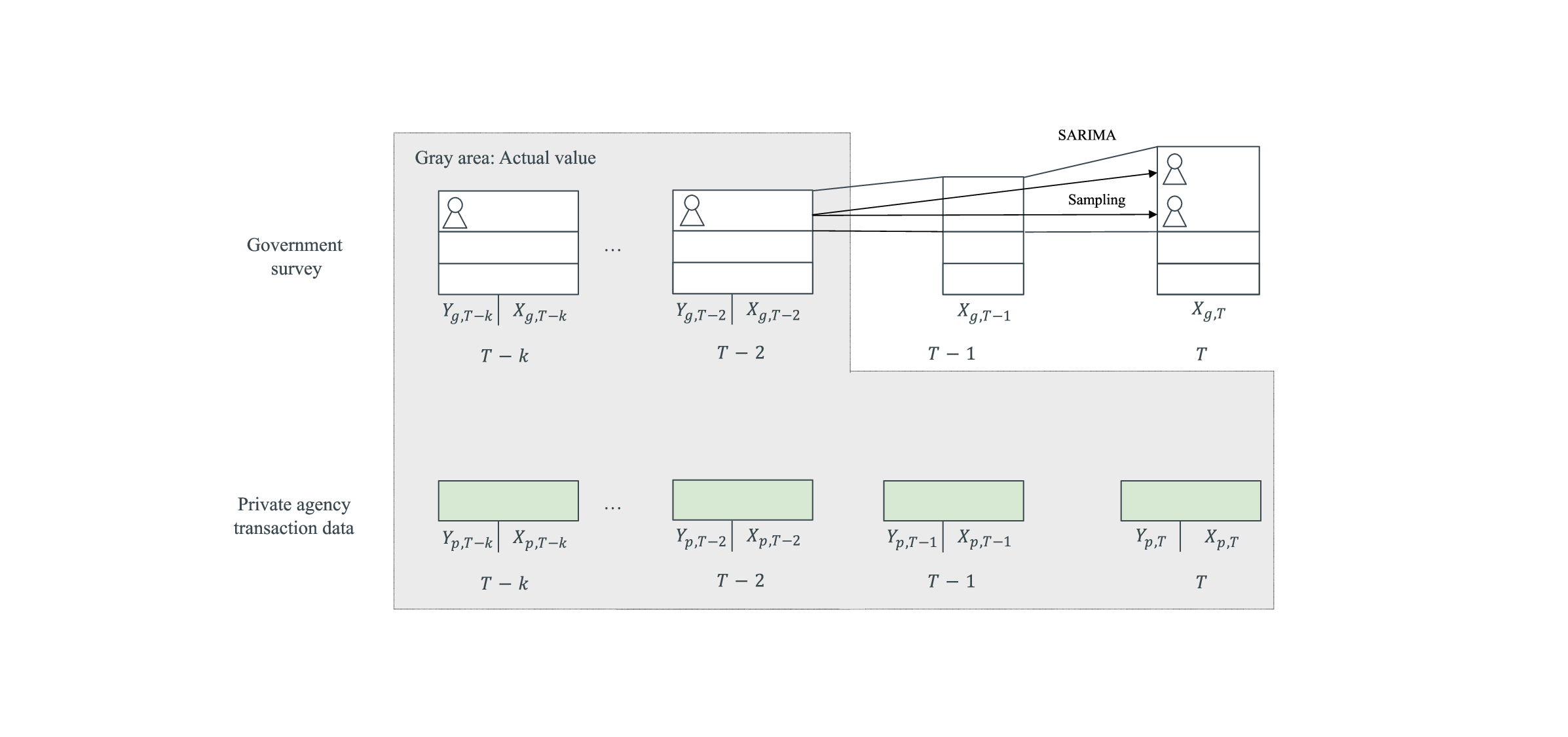}
\caption{Estimating the Number of Samples and Their Attributes Through SARIMA}\label{Fig5}
\end{figure*}

First, we determined the number of samples for the target half-year by various career-changing channels and gathered each sample's attribute information, as shown in Fig. \ref{Fig5}. In this phase, we did not obtain the label information for each sample in the target half-year. To estimate the number of samples for the target half-year by career-changing channels, we employed the SARIMA model using only the time series data of sample counts for each channel. The channels include public employment security offices, private employment agencies, job advertisements, personal connections, etc. Parameters were selected based on the Akaike Information Criterion (AIC). To acquire each sample’s attribute information, a duplicate random sampling was carried out from the latest value in the period for which there was available real data, i.e. the corresponding period one year ago.

Let $Y_{g,t}$ denote a vector of label information from the government survey samples. In other words, the components of this vector are the label information for each sample. The number of dimensions of this vector matches the number of samples. Subscript $t$ represents a time period, $T$ is the target period, and $k$ is the length of the training data. Let $X_{g,t}$ be a matrix of attribute information from the government survey samples. In other words, the column components of this matrix are the attribute information for each sample. The number of rows in this vector matches the number of samples. In this experiment, label information indicate whether the sample experienced a wage increase of over $10\%$ after changing careers. Time period $t$ represents a half-year period. Attribute information is made up of age, gender, and highest level of education, etc. The list of items used is shown in Section \ref{sec52}. $Y_{p,t}$ is a vector of label information from private employment agency samples, and $X_{p,t}$ is a matrix of transaction data representing attribute information of private employment agency samples.

In this process, we estimated the number of samples of $T$ in the government survey using SARIMA with the number of samples of $T-k$ to $T-2$ and obtain sample's attribute information in $T$ from $X_{g,T-2}$ by a duplicate random sampling. We did not estimate $Y_{g,T-1}$, $Y_{g,T}$, and did not use $Y_{p}$, or $X_{p}$ in this phase.

As mentioned above, the SARIMA estimates are made for each career change channel, but for one of these channels, via public job placement, it is possible to know the trend the number of people who decide to change jobs using another set of government-published preliminary statistics. As this study is intended for a practical task, we show how such preliminary government statistics can also be used as supplementary indicators. However, the improvement in accuracy resulting from this is minor and is not the essence of this study. The number of people who decide to change jobs via public job placement is published on a monthly basis. As shown in Fig. \ref{Fig6} below, up to one month in advance can be obtained at the time when estimating should be carried out. Therefore, the sample size via public job placement is not determined by the SARIMA method described above, but by an estimation based on the following formula.

\begin{equation}\label{SARIMA}
\widehat{{N_{g,T}}} = {N_{g,T-2}} \dfrac{n_{T}}{n_{T-2}}, 
\end{equation}

where $\widehat{{N_{g,T}}}$ is sample size of the government survey samples via public job placement for the target half year period $T$ and ${N_{g,T-2}}$ is the actual value for $T-2$ period. 
$n_{T}$ and $n_{T-2}$ represent the sample sizes indicated by the auxiliary series for the half-year periods $T$ and $T-2$. This is the five-month average of the auxiliary series on a monthly basis. For the first half of the year, it is the average from January to May, and for the second half of the year, it is the average from July to November.

\begin{figure*}[htbp]
\centering
\includegraphics[width=0.9\textwidth]{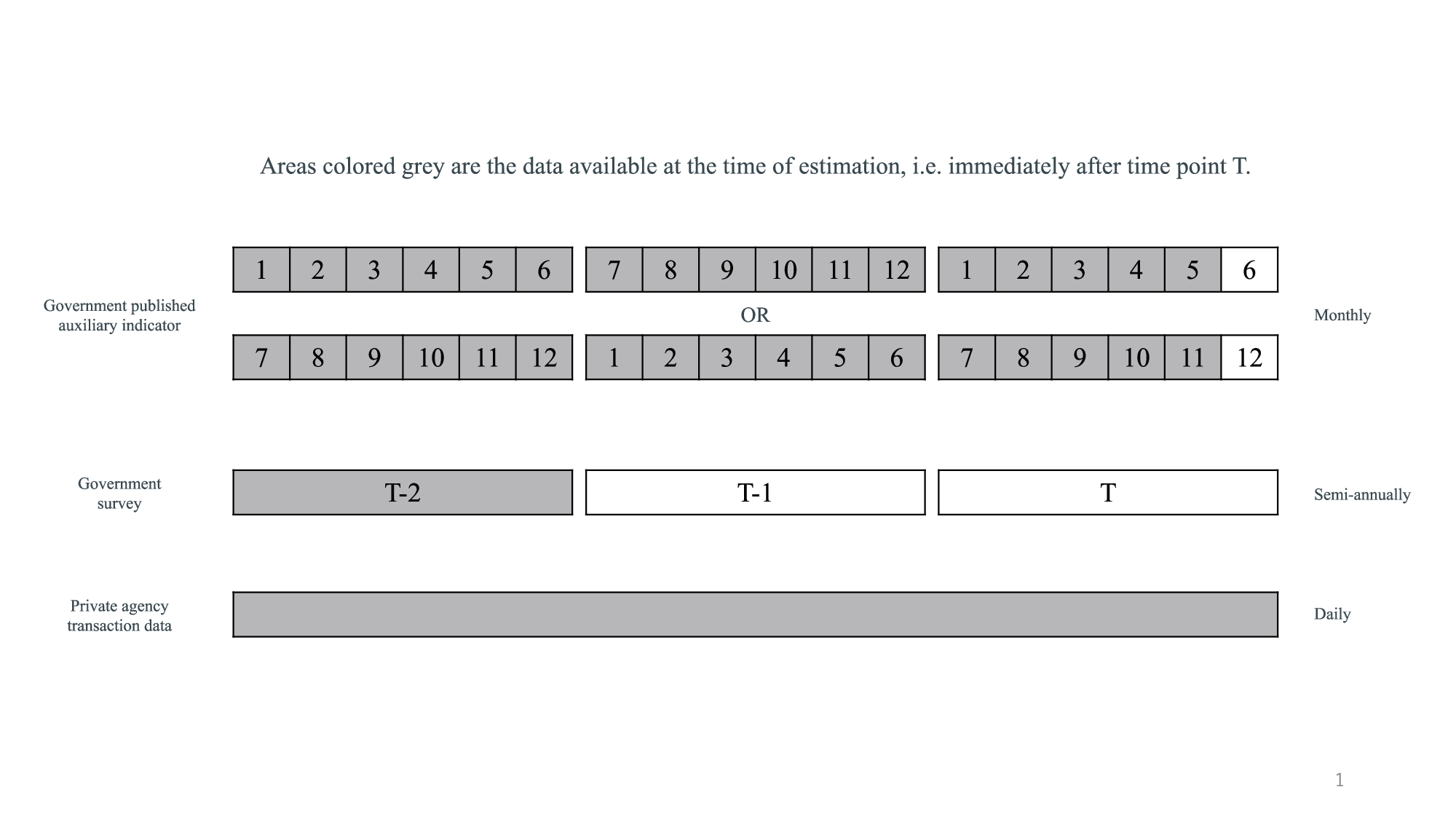}
\caption{Timing of when each piece of data can be obtained}\label{Fig6}
\end{figure*}

\subsection{Applying Density Ratio Estimation to Weight Private Agency Samples}\label{sec43}
\begin{figure*}[htbp]
\centering
\includegraphics[width=0.9\textwidth]{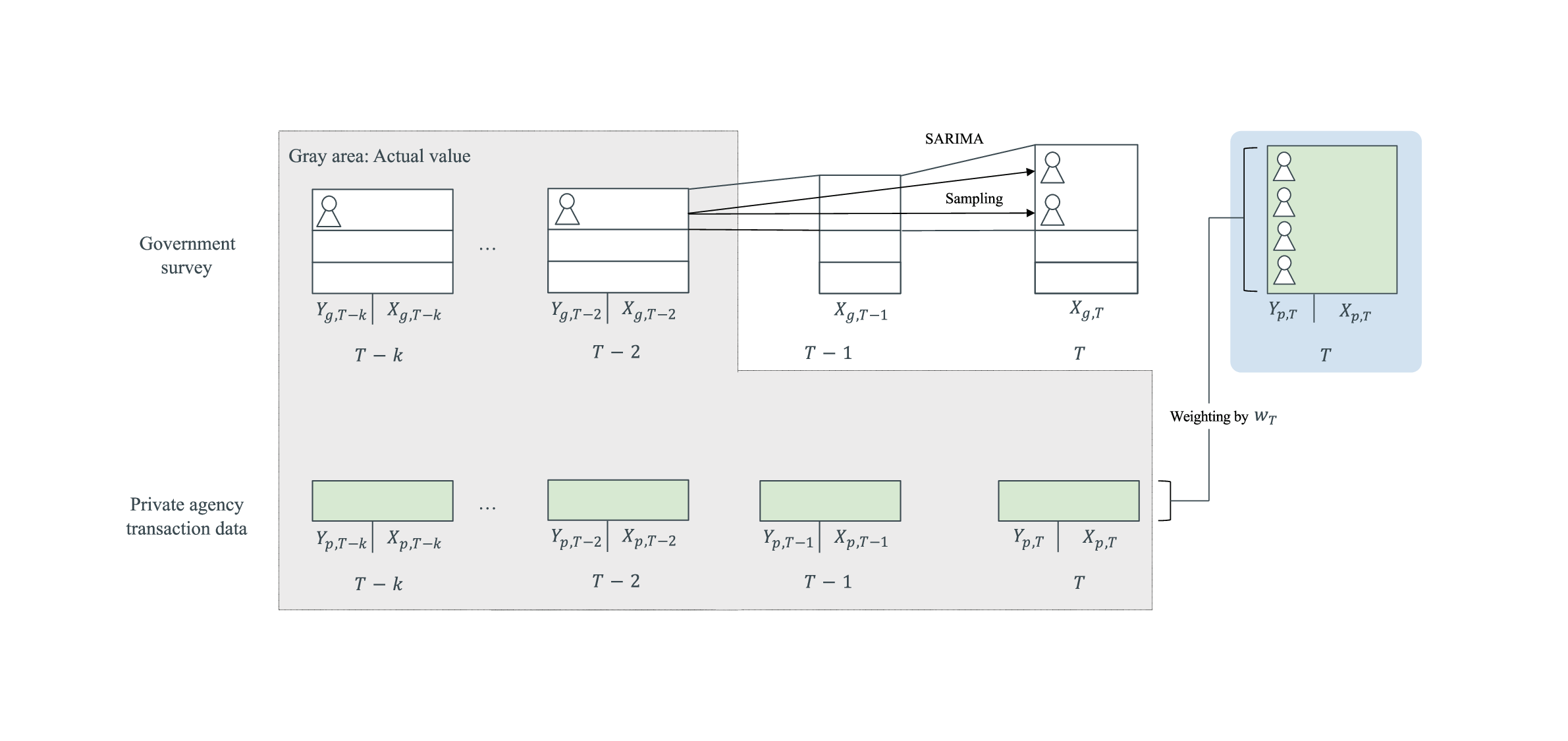}
\caption{Applying Density Ratio Estimation to Weight Private Agency Samples}\label{Fig7}
\end{figure*}

Next, we employed density ratio estimation to obtain the weight $w_{T}$, as shown in Fig. \ref{Fig7}, using $Y_{p,T}$, $X_{p,T}$, and $X_{g,T}$ from Section \ref{sec42}. The density ratio estimation problem is defined as follows \cite{bib17}. Let $\mathcal{D} \subset \mathbb{R}^{d}$ be the data domain, where $d$ is a positive integer. Suppose we have i.i.d. samples $\{x_{i}\}_{i = 1}^{n}$ from a distribution with density $p_{S}(x) > 0$ for all $x \in \mathcal{D}$, and i.i.d. samples $\{x_{j}^{'}\}_{j = 1}^{n^{'}}$ from another distribution with density $p_{T}(x) > 0$ for all $x \in \mathcal{D}$. The aim is to estimate the density ratio $w(x) = p_{T}(x) / p_{S}(x)$ from the samples $p_{T}(x)$ and $p_{S}(x)$.

A straightforward approach to approximating the density ratio is to estimate the numerator and denominator densities separately—typically by employing Kernel Density Estimation (KDE) \cite{bib34}—and then take their ratio. However, this approach is known to be unreliable in high-dimensional settings, as dividing two estimated quantities often amplifies the estimation error, particularly when the denominator is small or poorly estimated.

To address these limitations, a variety of methods have been developed to estimate the density ratio directly without explicitly estimating the individual densities. Representative approaches include moment matching methods, probabilistic classification-based methods, density matching techniques, and direct density ratio fitting methods.

Kernel Mean Matching \cite{bib35}, which is one of the moment matching methods, has a limitation in model selection. No known method can determine kernel parameters such as the width of a Gaussian kernel.

In the application of logistic regression \cite{bib21, bib22, bib36}, which is a probabilistic classification-based method, and the Kullback-Leibler Importance Estimation Procedure(KLIEP) \cite{bib37}, which is a density matching technique, cross-validation can be used to optimize the tuning parameters. However, this process is time-consuming because a non-linear optimization problem must be solved.

Least squares importance fitting (LSIF) and unconstrained LSIF (uLSIF) \cite{bib25} represent direct density ratio fitting methods. In LSIF, cross-validation can be used to optimize the tuning parameters. LSIF is more computationally efficient than the application of logistic regression and KLIEP, but it tends to be numerically unstable. In contrast, uLSIF is both fast and reliable as it can be computed analytically. This method is considered the most practical in real-world applications. Therefore, we selected uLSIF. To introduce uLSIF, we begin by providing an overview of LSIF.

\subsubsection{Least-Squares Importance Fitting (LSIF) }\label{sec431}
The fundamental concept of LSIF is to transform the density ratio estimation problem into a least-squares function fitting problem \cite{bib25}. The density ratio $w(x)$ is modeled by the following linear model:

\begin{equation}\label{e1}
\hat{w}(x)= \sum_{l=1}^{b} \alpha_{l} \phi_{l}(x) 
= \boldsymbol{\phi(x)}^{\mathsf{T}} \boldsymbol{\alpha}, 
\end{equation}

where $\boldsymbol{\alpha} = (\alpha_{1}, \alpha_{2}, ..., \alpha_{b})^{\mathsf{T}} $is a parameter vector and $\boldsymbol{\phi(x)}: \mathbb{R}^{d} \rightarrow \mathbb{R}^{b}$ is a non-negative basis function vector.
$\boldsymbol{\alpha}$ is decided so that the following squared error $J_{0}$ is minimized:

\begin{equation}\label{e2}
\begin{split}
J_{0}(\boldsymbol{\alpha})  := &\dfrac{1}{2} \int ( \hat{w}(x) - w(x) )^{2} p_{S}(x) dx \\
= &\dfrac{1}{2} \int \hat{w}(x)^{2} p_{S}(x) dx \\
& - \int \hat{w}(x)^{2} p_{T}(x) dx + \dfrac{1}{2} \int w(x) p_{T}(x) dx.
\end{split}
\end{equation}

We can safely ignore the third term on the right hand of the (\ref{e2}) because it is a constant.
Let us denote the first two terms on the right hand of the (\ref{e2}) by $J$.

\begin{equation}\label{e3}
J(\boldsymbol{\alpha})  := \dfrac{1}{2} \int \hat{w}(x)^{2} p_{S}(x) dx - \int \hat{w}(x)^{2} p_{T}(x) dx 
\end{equation}

Approximating the expectations in $J$ by empirical averages, we obtain
\begin{equation}\label{e4}
\begin{split}
\hat{J}(\boldsymbol{\alpha})  := &\dfrac{1}{2n} \sum_{i=1}^{n} \hat{w}(x_{i})^{2} 
                                                           - \dfrac{1}{n^{'}} \sum_{j=1}^{n^{'}} - \hat{w}(x_{j}^{'}) ^{2}  \\
                                                     = &\dfrac{1}{2} \sum_{l,l^{'}=1}^{b} \alpha_{l} \alpha_{l^{'}}  \hat{H}_{l,l^{'}}
                                                           -  \sum_{l = 1}^{b} \alpha_{l}   \hat{h}_{l} ,                                     
\end{split}
\end{equation}

where 
\begin{equation}\label{e5}
\begin{split}
\hat{H}_{l,l^{'}} := &\dfrac{1}{n} \sum_{i=1}^{n} \phi_{l}(x_{i}) \phi_{l^{'}}(x_{i}),  \\
\hat{h}_{l} := &\dfrac{1}{n^{'}} \sum_{i=1}^{n^{'}} \phi_{l}(x_{i}^{'}).    
\end{split}
\end{equation}

Then, the optimization problem is expressed as

\begin{equation}\label{e6}
\begin{split}
\underset{\{\alpha_{l}\}_{l = 1}^{b}}{\min} \left [
\dfrac{1}{2} \sum_{l,l^{'}=1}^{b} \alpha_{l} \alpha_{l^{'}}  \hat{H}_{l,l^{'}}
-  \sum_{l = 1}^{b} \alpha_{l}   \hat{h}_{l} 
-  \lambda \sum_{l = 1}^{b} \alpha_{l} \right ] \\
subject \: to \: \alpha_{1}, \alpha_{2}, ..., \alpha_{b} 	\geq 0, 
\end{split}
\end{equation}

where $\lambda$ is the non-negative regularization parameter.
This approach is known as LSIF or constrained LSIF.

\subsubsection{unconstrained LSIF (uLSIF) }\label{sec432}
Next, we explain the implementation of LSIF without applying the non-negativity constraint. In the absence of the non-negativity constraint, the regularizer in (\ref{e6}) becomes ineffective. Hence, a quadratic regularizer is employed \cite{bib25}. This leads us to the following optimization problem:

\begin{equation}\label{e7}
\underset{\{\alpha_{l}\}_{l = 1}^{b}}{\min} \left [
\dfrac{1}{2} \sum_{l,l^{'}=1}^{b} \alpha_{l} \alpha_{l^{'}}  \hat{H}_{l,l^{'}}
-  \sum_{l = 1}^{b} \alpha_{l}   \hat{h}_{l} 
-  \dfrac{\lambda}{2}  \sum_{l = 1}^{b} \alpha_{l}^{2} \right ]. 
\end{equation}

The solution to equation (\ref{e7}) can be determined analytically by applying the following equations:

\begin{equation}\label{e8}
\begin{split}
\tilde{\boldsymbol{\alpha} } =  (\tilde{\alpha_{1}}, \tilde{\alpha_{2}}, ..., \tilde{\alpha_{b}})^{\mathsf{T}} &
=  ( \boldsymbol{\hat{H}} + \lambda \boldsymbol{I_{b}})^{-1} \boldsymbol{\hat{h}} \\
\boldsymbol{\hat{H}} := &\dfrac{1}{n} \sum_{i=1}^{n} \boldsymbol{\phi(x_{i})} \boldsymbol{\phi(x_{i})^{  \mathsf{T}  }}\\ 
\boldsymbol{\hat{h}} := & \dfrac{1}{n^{'}} \sum_{j=1}^{n^{'}} \boldsymbol{\phi(x_{j}^{'})} ,
\end{split}
\end{equation}

where $\boldsymbol{I_{b}}$ is the b-dimensional identity matrix.
Since the non-negativity constraint was removed, the estimated density ratio values could be negative. To address this, negative values can be rounded up to zero as follows:

\begin{equation}\label{e9}
\tilde{\alpha_{l}} = \max(0, \tilde{\alpha_{l}}) \: for \: l = 1,2,...,b.
\end{equation}
This method is called an unconstrained LSIF: uLSIF.\\

Following the approach in \cite{bib25}, we employ the Gaussian kernel as a basis function as follows:

\begin{equation}\label{e10}
K_{\sigma}(x,x^{'}) := \exp \left(  \dfrac{\| x-x^{'} \|^{2}}{2\sigma^{2}}  \right).
\end{equation}

\subsection{Supervised Learning under Covariate Shift: Classification}\label{sec44}
Next, we carried out supervised learning under covariate shift using the weighted samples from private employment agencies obtained in Section \ref{sec43}. In this phase, we attempted both classification and regression. In Section \ref{sec44}, we explain the classification. The regression is discussed in Section \ref{sec45}.
We represented $X_{g,t}$ as a vector of attribute information from government survey samples, $Y_{p,t}$ as a vector of label information from private employment agency samples, $X_{p,t}$ as a vector of attribute information from private employment agency samples, and $w_{t}$ as the weight for private employment agency samples calculated in Section \ref{sec43}.

\begin{figure*}[htbp]
\centering
\includegraphics[width=0.9\textwidth]{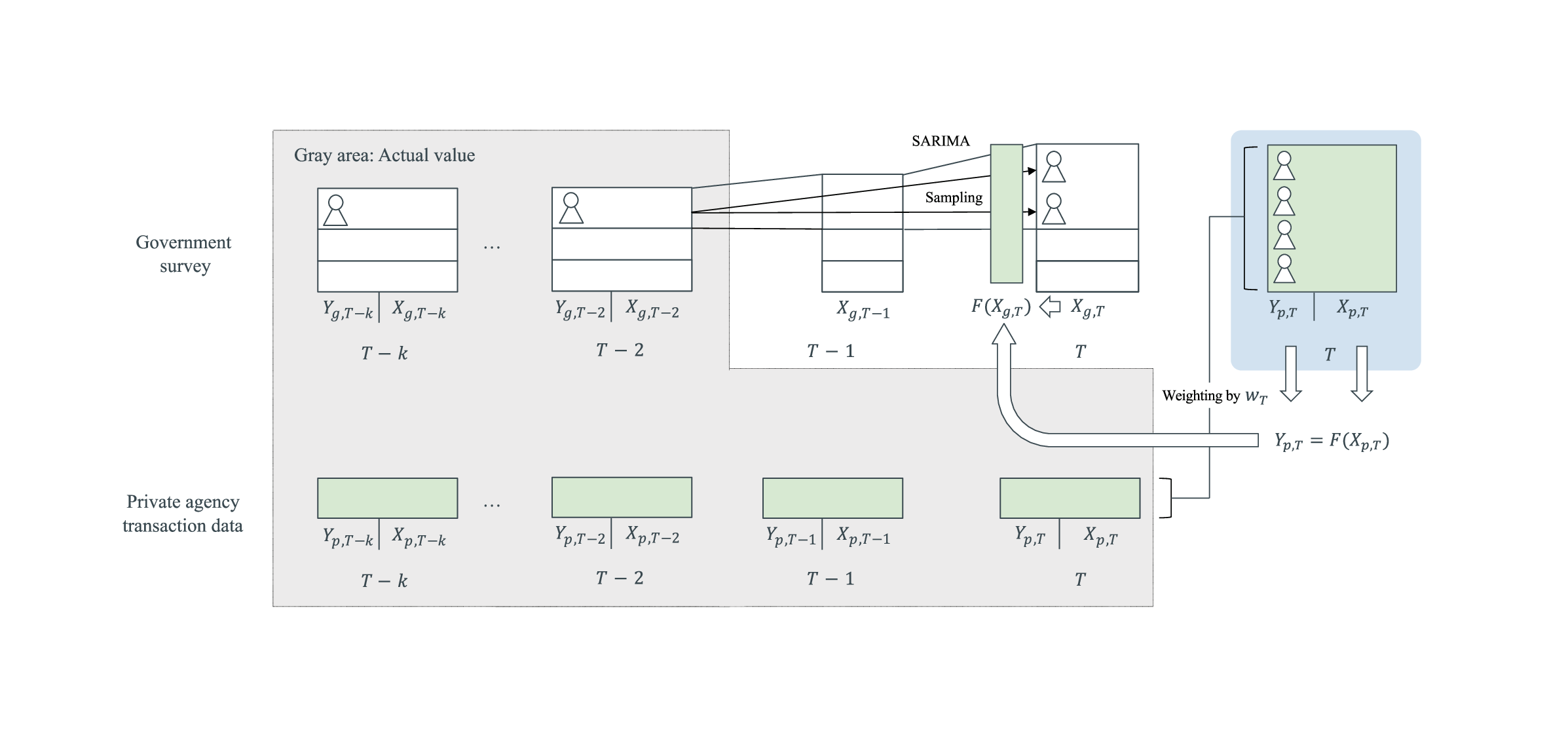}
\caption{Supervised Learning under Covariate Shift}\label{Fig8}
\end{figure*}

Here, we estimated $F$ such that $Y_{p,T} = F(X_{p,T})$ with the weight $w_{T}$, as illustrated in Fig. \ref{Fig8}. We compared logistic regression models with elastic net penalties, random forest classifiers, and gradient-boosted decision tree classifiers as $F$. All attribute information described in Section \ref{sec52} was used as explanatory variables $X_{p,t}$.
Covariate shift refers to a situation where the training and test input points follow different probability distributions, but the conditional distributions of output values given input points remain unchanged, a scenario often encountered in machine learning \cite{bib17}. According to the notation in Section \ref{sec43}, let $x \in \mathcal{D} \subset \mathbb{R}^{d}$ and $y \in \mathcal{D^{'}} \subset \mathbb{R}$ represent the covariate and its class label, respectively. The situation can be described as follows:

\begin{equation}\label{covariate shift2} 
p_{S}(y \mid x) = p_{T}(y \mid x)  \; \; \text{and} \; \;  p_{S}(x) \neq p_{T}(x).
\end{equation}

Standard learning methods like maximum likelihood estimation are biased under covariate shift. However, we can asymptotically correct this bias by weighting the loss function based on the density ratio \cite{bib28}.

\subsection{Supervised Learning under Covariate Shift: Regression}\label{sec45}
Here, we discuss regression in supervised learning under covariate shift. Unlike the categorical label information in government survey samples, the label information from private employment agency samples is continuous. To leverage this continuous data, we performed regression analysis. In Section \ref{sec44}, a label was $1$ if the sample had a wage increase of over $10\%$ after changing careers, and $0$ if not. Here in Section \ref{sec45}, the label information is continuous. The numerator of the label is the wage after changing careers, and the denominator is the wage before changing careers. If a sample had a $10\%$ wage increase after changing careers, the label is $1.1$. After obtaining $F(X_{g,T})$, we convert this value to a classifier score ranging from $0$ to $1$. We compared linear regression with an elastic net penalty, the Random Forests regressor, and the Gradient-Boosted Decision Trees regressor as $F$. All attribute information listed in Section \ref{sec52} was used as explanatory variables $X_{p,T}$. $Y_{p,t}$ was pre-transformed using Box-Cox transformation. The conversion to a classifier score was done by calculating the probability that the value of $F(X_{p,T})$ exceeds the Box-Cox transformed value of $1.1$, assuming the residuals follow a normal distribution. For example, if the value of $F(X_{p,T})$ equals the Box-Cox transformed value of $1.1$, the converted score is $0.5$.

\subsection{Correction of the Label Information}\label{sec46}
In Section \ref{sec43}, Section \ref{sec44}, or the corresponding value in Section \ref{sec45}, we obtained bias-corrected label information for the target half-year.
We can then calculate the proportion of individuals whose wages increased after changing careers. However, this time, we did not use these simple calculation results. Instead, we used results calculated from bias-corrected label information for the target half-year, corrected by the method described below.

\subsubsection{Classification of Selection Bias and Positioning of the Case}\label{sec461}
Again, as discussed in Section 3, covariate shifts can be regarded as a particular class of selection bias. Since \cite{bib29}, it has become customary to handle sample selection bias by dividing missing data mechanisms into three categories: MCAR, MAR, and NMAR.

\begin{itemize}
\item MCAR: Missing completely at random refers to a situation where the probability of data being missing is unrelated to the specific value that should be obtained or the set of observed responses.

\item MAR: Missing at random is a more realistic condition where the probability of responses being missing depends on the set of observed responses but not on the specific missing values. 

\item NMAR: Not missing at random is a more challenging situation where the missing data pattern is non-random and depends on the missing variables.
\end{itemize}

According to \cite{bib30}, the covariate shift assumption is considered equivalent to the MAR assumption. However, as \cite{bib31} points out, since MAR and NMAR are not fundamentally distinct but rather exist on a continuum, we regard the situation addressed in this study as more appropriately classified as NMAR.

In NMAR, as stated in \cite{bib32}, the response probability cannot be verified using only the observed study variables, and therefore additional assumptions are often required. The correction method outlined below corresponds to this additional assumption.

\subsubsection{Additional Assumption}\label{sec462}
As discussed earlier, covariate shift is a condition where training and test input points follow different probability distributions, but the conditional distributions of output values given input points do not change.

According to the notation in Section \ref{sec43}, let $x \in \mathcal{D} \subset \mathbb{R}^{d}$ and $y \in \mathcal{D^{'}} \subset \mathbb{R}$ represent the covariate and its class label, respectively. The situation is expressed as:

\begin{equation}\label{covariate shift3}
p_{S}(y \mid x) = p_{T}(y \mid x) \; \; \text{and} \; \;  p_{S}(x) \neq p_{T}(x).
\end{equation}

In this context, the attribute information of private employment agency samples and government survey samples follows different probability distributions, but the conditional distributions of label values given attribute information remain unchanged. Label information indicates whether individuals had a wage increase of more than $10\%$ after changing careers. However, we did not obtain enough attribute information to assume the above situation. Therefore, we assumed the following scenario:

\begin{equation}\label{covariate shift beta}
\beta p_{S}(y = 1 \mid x) = p_{T}(y = 1 \mid x) \; \;  \text{and} \;  \; p_{S}(x) \neq p_{T}(x),
\end{equation}

where $\beta$ is a constant. This assumption means that the bias that cannot be corrected by attribute information is constant. The precise calculation method for $\beta$ is described in Section \ref{sec53}.

We multiplied this $\beta$ by the classifier score estimated in Section \ref{sec43}, Section \ref{sec44}, or the score corresponding to the classifier score estimated in Section \ref{sec45}. The average of these corrected scores was considered in the estimated value of wage changes for hired career-changing employees, representing the proportion of individuals with increased wages after changing careers. We had the option to derive the estimated value by converting the score to $1/0$ using a cut-off point and counting the proportion of $1$. However, due to the difficulty of setting appropriate cut-off points, we did not choose this option.

\section{Experiment}\label{sec5}
While Section \ref{sec3} maintained a certain level of generality, providing a somewhat abstract description of the task setting, this section presents the details of the experimental design that were not discussed there. Specifically, we describe the time lag structure of the official statistics used in the experiment, the data characteristics, and the details of experimental design.

\subsection{Time Lag Structure}\label{sec51}
\begin{figure*}[htbp]
\centering
\includegraphics[width=1.5\textwidth]{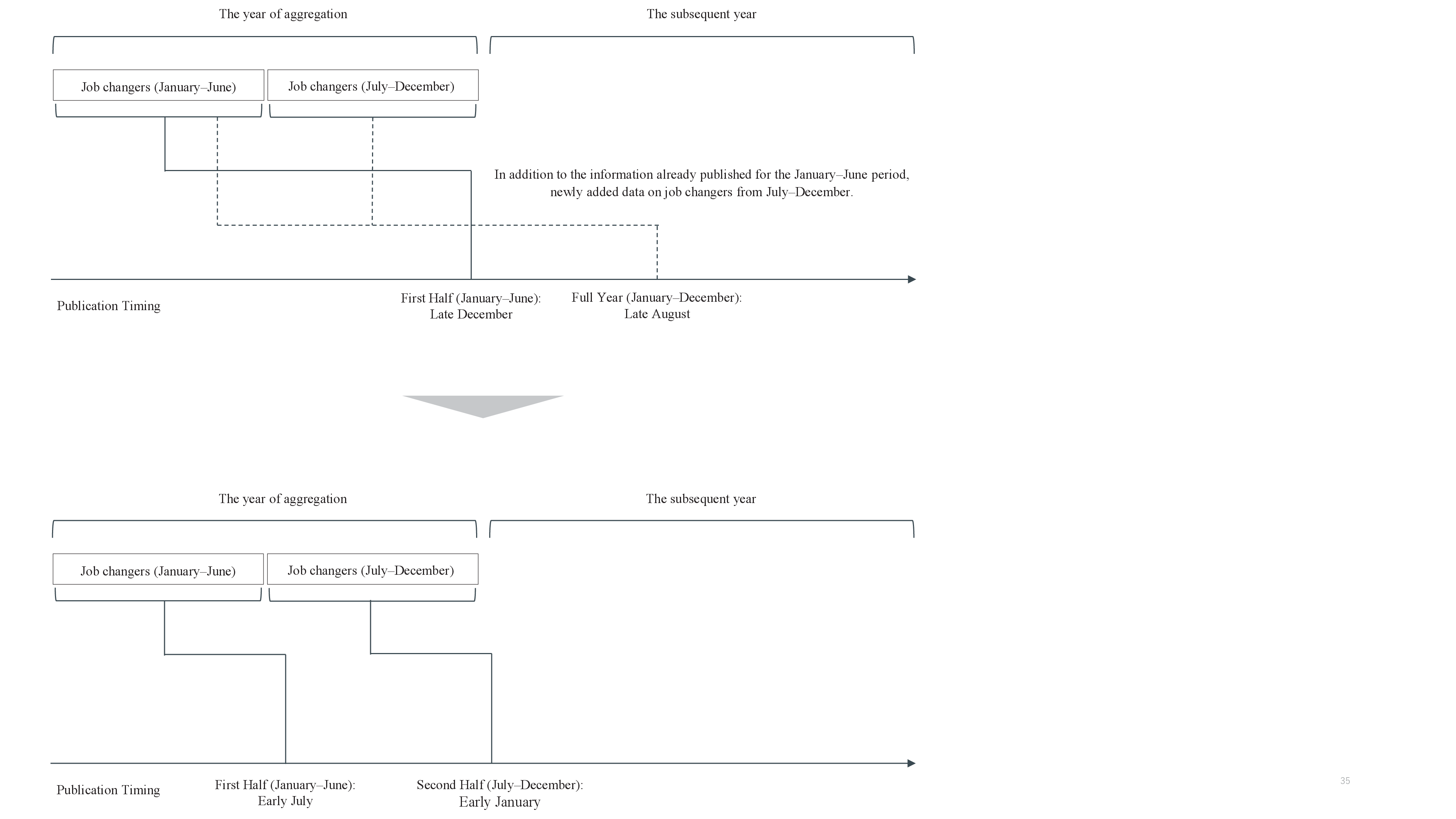}
\caption{Publication Timing}
\label{Fig9}
\end{figure*}

The official statistics used in the experiment are disseminated through a semiannual statistical survey, experiencing a publication delay of six to thirteen months. As illustrated in Fig. \ref{Fig9}, data from January to June are released at the end of December, while data for July to December are made public in late August of the subsequent year. This temporal delay precludes policymakers and hiring managers from utilizing the indicator for timely decision-making.

To address this issue, we utilized transaction data from a private employment agency, specifically Recruit Agent, Japan’s largest such agency under the Recruit Holdings Co., Ltd. umbrella, which provides data with virtually no delay. We are able to access data from the previous day on the current day. Consequently, our objective is to expedite the readiness of the January-June period data for release in early July and the July-December period data in early January, by leveraging this novel data source. This approach effectively reduces the time lag from several months to merely a few days.

In this study, both the sample of job changers obtained through government surveys and the sample of job changers obtained from private recruitment transaction data are used for the estimation. The sample of job changers obtained from government surveys is available within the government immediately after the survey is completed, whereas outside the government it is available after the publication of the statistical indicator. We consider a realistic setting in which the estimation is conducted by an organization external to the government. Letting $T$ denote a six-month period, we assume that when the sample for period $T$ from private recruitment transaction data becomes available, the government survey sample is only available up to period $T-2$, as illustrated in Fig. \ref{Fig5} through Fig. \ref{Fig8} in Section \ref{sec4}.

\subsection{Data Characteristics}\label{sec52}
In this study, we utilize two data sources: samples from a government survey and private employment agency samples.

First, we detail the government survey samples used.
These samples were collected by the Ministry of Health, Labor, and Welfare in Japan through a sample survey called the Survey on Employment Trends. While only statistical information is publicly available and the samples and weight information are not published, we received these from the Ministry.
Though the survey targets companies, it also includes information on individuals who joined the company, and we used individual unit samples and weight information for this period. In Section \ref{sec4}, $Y_{g,t}$ and $X_{g,t}$ denote the vector of the government survey sample's label and attribute information. The samples were replicated according to the weight information proportions.
The survey is conducted semiannually. We used data from the first half-year of 2004 to the first half-year of 2018. The number of new employee samples in the first half of 2018 is 37,841. The estimated number of job changers from the samples and weight information for the first half of 2018, published by the government in the Survey on Employment Trends, is roughly 2,671,100.
Of these, approximately 1,651,400 are full-time employees and 1,019,800 are part-time workers. The data items for the samples are listed in below.
We focused on the statistics and samples of full-time employees here.
Therefore, samples of part-time workers were excluded.

\begin{itemize}
  \item \textbf{Filter information:}
    \begin{itemize}
      \item Hiring route
      \item Type of employment after changing careers
      \item Type of employment before changing careers
    \end{itemize}
  \item \textbf{Attribute information:}
    \begin{itemize}
      \item Age
      \item Gender
      \item Highest level of education
      \item Location of the company the sample belongs/ed to after changing careers
      \item Industry of the company the sample belongs/ed to after changing careers
      \item Number of employees of the company the sample belongs/ed to after changing careers
      \item Location of the company the sample belonged to before changing careers
      \item Industry of the company the sample belonged to before changing careers
      \item Number of employees of the company the sample belonged to before changing careers
    \end{itemize}
  \item \textbf{Label information:}
    \begin{itemize}
      \item Whether the sample had over 10\% increased wage after changing careers
      \item Wage before/after job change (only in private employment agency samples)
    \end{itemize}
\end{itemize}

Second, we provide an explanation of the private employment agency samples.
These samples are sourced from the transaction data of the service Recruit Agent of the private employment agency under Recruit Holdings Co., Ltd.
The agency currently registers over 1,200,000 new job seekers annually and has more than 500,000 job offers. At present, about 50,000 individuals change jobs through this agency each year. Job seeker’s attribute information, detailed job offer information, and information about the job search process, including applications and interviews, are recorded.
We processed the transaction data to conform to the data items of the government survey presented above. In addition, we have extra items as label information.
While we can only determine whether the sample had over $10\%$ increased wage after changing careers as categorical data in the government survey samples, the transaction data includes wages before and after changing careers as continuous values. This continuous label information was used for regression, as explained in Section \ref{sec45}.

The transaction data is available in real time, but the coverage rate is quite low - less than $5\%$ even for the largest agency case used in this study - and furthermore, when considered as a sample to understand the whole of Japan, it has a very strong bias. As an illustrative example, in the second half of 2018, the average age of the population in the government survey—namely, full-time employees who changed jobs—was approximately 40 years. In contrast, the corresponding figure in the transaction data was around 31 years. Similarly, the proportion of individuals with a university degree or higher was about $35\%$ in the official statistics, whereas it was approximately $79\%$ in the transaction data, indicating a substantial discrepancy between the two datasets.

\subsection{Details of Experimental Design}\label{sec53}
To measure the performance of our estimation, we used MAE, as discussed in Section \ref{sec32}. We calculated MAE by comparing our estimated values of the proportion of individuals with increased wages after changing careers to actual values over the validation period. We used the dataset from the first half-year of 2004 to the first half-year of 2018. The validation period spans from the first half-year of 2013 to the first half-year of 2018.

We estimated the target half-year in the validation period using government survey samples up to a year before the target half-year and private employment agency samples from the target half-year.

For example, as mentioned in Section \ref{sec42}, we used government survey samples from the first half-year of 2004 to the second half-year of 2013 to estimate the second half-year of 2014 using SARIMA. Next, as mentioned in Section \ref{sec43}, we used private employment agency samples from the second half-year of 2014 for density ratio estimation using uLSIF.

In Section \ref{sec43}, we conducted density ratio estimation with uLSIF. We tried several variable combinations for input and showed two cases in Section \ref{sec6}. In the first case, we used all data items described in Section \ref{sec52}. In the second case, we used three items: age, highest level of education, and number of employees of the company the sample belonged to before changing careers.

In Section \ref{sec46}, we corrected the label information with a constant value $\beta$. $\beta$ was calculated as the ratio of two values.

The numerator was the actual value of the proportion of individuals with increased wages after changing careers. The denominator was the estimated value without this correction. We tried two cases to calculate $\beta$:
In the first case, we calculated $\beta$ as the average of all estimation results from the first half-year of 2013 to the second half-year of 2018. In this case, we estimated with data that were unavailable then. In other words, the estimation assumes that $\beta$ is a time-independent parameter identified by the method discussed above. If this assumption is not satisfied, this evaluation is inappropriate. In the second case, we calculated $\beta$ as the average of estimation results from the first half-year of 2013 to the last half-year before the target half-year. This means we estimated with the data available at that time.

\section{Results}\label{sec6}
Table \ref{Tab1} and Table \ref{Tab2} presents the MAE for the validation period spanning from the second half of 2013 to the second half of 2018.

In Section \ref{sec46}, we adjusted the label information using constant values denoted by $\beta$. To determine $\beta$, we examined two scenarios: taking the average of all estimation results and taking the average of estimation results obtained prior to the target period.

Table \ref{Tab1} shows the case we calculated $\beta$ as the average of estimation results from the first half-year of 2013 to the last half-year before the target half-year. This means we estimated with the data available at that time.

Table \ref{Tab2} shows the case we calculated $\beta$ as the average of all estimation results from the first half-year of 2013 to the second half-year of 2018. In this case, we estimated with data that was unavailable then. In other words, the estimation assumes that $\beta$ is a time-independent parameter identified by the method discussed above. If this assumption is not satisfied, this evaluation is inappropriate.

In Section \ref{sec43}, we utilized uLSIF for density ratio estimation with various combinations of input variables. Here, we illustrate two representative cases: the use of all items and the use of a selected subset consisting of three items.

As discussed in Section 3.2, we aim to evaluate the extent to which accuracy improves through the mitigation of selection bias. Therefore, the simple extrapolation-based prediction(Simple extrapolation) shown in Equation \ref{Simple} is used as a benchmark.

As explained in Section \ref{sec41}, we implemented two approaches: one that applies only weighting based on density ratios(Weighting only), as shown on the left side of Fig. \ref{Fig4} in Section \ref{sec41}, and another that includes an additional supervised learning step, as illustrated on the right side of the same figure. As discussed in Section \ref{sec44} and Section \ref{sec45}, we explored both classification(Cls) and regression(Reg) approaches for supervised learning under covariate shift. For the classification task, we compared the logistic regression model with an elastic net penalty(EN), the random forest classifier(RF), and the gradient-boosted decision trees classifier(GB). For the regression task, we compared the linear regression model with an elastic net penalty(EN), the random forest regressor(RF), and the gradient-boosted decision trees regressor(GB).

The abbreviations shown in parentheses correspond to the notations used within Table \ref{Tab1} and Table \ref{Tab2}.

\begin{table*}[h]
\centering
\caption{MAE in the case of calculating $\beta$ from the results before the target period}\label{Tab1}

\begin{minipage}{0.45\textwidth}
\centering
\begin{tabular}{@{}ll@{}}
\hline
\multicolumn{2}{c}{Simple extrapolation} \\
\hline
& 2.55 \\
\hline
\\
\hline
\multicolumn{2}{c}{Weighting only} \\
\hline
uLSIF with & \\
three items & 1.85 \\
all items   & 1.82 \\
\hline
\end{tabular}
\end{minipage}
\hfill
\begin{minipage}{0.52\textwidth}
\centering
\vspace{2ex}
\begin{tabular}{@{}lll@{}}
\hline
\multicolumn{3}{c}{+Supervised learning} \\
\hline
\multicolumn{3}{c}{EN} \\
\hline
uLSIF with & Cls & Reg \\
three items & 1.92 & 1.72 \\
all items   & 1.74 & 1.77 \\
\hline
\multicolumn{3}{c}{RF} \\
\hline
uLSIF with & Cls & Reg \\
three items & 1.74 & 1.72 \\
all items   & 1.60 & 1.70 \\
\hline
\multicolumn{3}{c}{GB} \\
\hline
uLSIF with & Cls & Reg \\
three items & 1.75 & 2.00 \\
all items   & 1.68 & 1.79 \\
\hline
\end{tabular}
\end{minipage}

\end{table*}

\begin{table*}[h]
\centering
\caption{MAE in the case of calculating $\beta$ from all results}\label{Tab2}

\begin{minipage}{0.45\textwidth}
\centering
\begin{tabular}{@{}ll@{}}
\hline
\multicolumn{2}{c}{Simple extrapolation} \\
\hline
& 2.55 \\
\hline
\\[-6pt]
\hline
\multicolumn{2}{c}{Weighting only} \\
\hline
uLSIF with & \\
three items & 1.62 \\
all items   & 1.67 \\
\hline
\end{tabular}
\end{minipage}
\hfill
\begin{minipage}{0.52\textwidth}
\centering
\vspace{2ex}
\begin{tabular}{@{}lll@{}}
\hline
\multicolumn{3}{c}{+Supervised learning} \\
\hline
\multicolumn{3}{c}{EN} \\
\hline
uLSIF with & Cls & Reg \\
three items & 1.57 & 1.54 \\
all items   & 1.59 & 1.63 \\
\hline
\multicolumn{3}{c}{RF} \\
\hline
uLSIF with & Cls & Reg \\
three items & 1.62 & 1.49 \\
all items   & 1.48 & 1.57 \\
\hline
\multicolumn{3}{c}{GB} \\
\hline
uLSIF with & Cls & Reg \\
three items & 1.63 & 1.82 \\
all items   & 1.46 & 1.57 \\
\hline
\end{tabular}
\end{minipage}

\end{table*}

In what follows, we systematically summarize the results.

The first point to note is that in both Tables \ref{Tab1} and Tables \ref{Tab2}, the Weighting only method clearly outperforms the Simple extrapolation method. This indicates that the approach illustrated on the left side of Fig. \ref{Fig4} in Section \ref{sec41} is sufficient to mitigate selection bias and can lead to a substantial improvement in predictive accuracy.

The next important observation is that the addition of a supervised learning step generally improves accuracy compared to the Weighting only method. In other words, the approach shown on the right side of Fig. \ref{Fig4} in Section \ref{sec41} demonstrates better performance than the approach on the left, highlighting the benefit of incorporating this additional step.

Of particular interest is the case where all items are used in uLSIF: in all 12 cases across the two tables, we observe improvements in accuracy when the supervised learning step is added to Weighting only. When only three items are used, improvements are still observed, but in only 7 out of the 12 cases. 
In the Weighting only case, the number of variables applied to uLSIF did not appear to influence performance; however, when supervised learning is employed, inaccurate weighting likely distorts the learned function, and the adverse effects of such distortion are amplified. However, using more variables does not necessarily lead to better results. In the present case, using all items rather than restricting to only three variables appears to have enabled more appropriate learning. Nevertheless, since density ratio estimation is prone to overfitting, caution is warranted—more variables are not always better.

With respect to model choice, there is a slight tendency for more flexible models to outperform linear models in terms of accuracy. In this study, we also explored regression models, given the availability of continuous-valued supervisory signals with richer information content as discussed in Section \ref{sec45}. However, their performance did not exceed that of classification models.

Regarding $\beta$, the case of calculating $\beta$ from all results generally showed better accuracy than the case of calculating $\beta$ from the results before the target period. This suggests that $\beta$ is relatively stable over time. It should be noted, however, that in practice, the case of calculating $\beta$ from all results is infeasible, as it relies on data that are not available at the time of estimation.

For reference, we tested whether these methods had better predictive power than the naive extrapolation method shown in Equation \ref{Simple} by conducting a HLN test. The results are shown in Table \ref{Tab3} and Table \ref{Tab4}. The abbreviations used in the tables are consistent with those in Tables \ref{Tab1} and \ref{Tab2}. $*$ indicates significance at the $5\%$ level, and $**$ indicates significance at the $1\%$ level. Because $\dag$ does not meet the significance level above, it cannot be considered to be a significant result. However, because it meets the $10\%$ significance level, it can be described as showing a trend toward significance. While not all patterns yielded significant results, a substantial number of cases exhibited statistical significance.

\begin{table*}[h]
\centering
\caption{HLN test results in the case of calculating $\beta$ from the results before the target period}\label{Tab3}

\begin{minipage}{0.45\textwidth}
\centering
\vspace{1ex}
\begin{tabular}{@{}lll@{}}
\hline
\multicolumn{3}{c}{uLSIF with three items} \\ 
\hline
Model & HLN stat & P-value \\
\hline
EN Cls & -0.956 & 0.181 \\
EN Reg & -1.078 & 0.153 \\
RF Cls & -1.829 & $0.049^{*}$ \\
RF Reg & -1.960 & $0.039^{*}$ \\
GB Cls & -1.950 & $0.040^{*}$ \\
GB Reg & -0.678 & 0.257 \\
\hline
\end{tabular}
\end{minipage}
\hfill
\begin{minipage}{0.52\textwidth}
\centering
\vspace{1ex}
\begin{tabular}{@{}lll@{}}
\hline
\multicolumn{3}{c}{uLSIF with all items} \\ 
\hline
Model & HLN stat & P-value \\
\hline
EN Cls & -1.692 & $0.061^{\dag}$ \\
EN Reg & -3.050 & $0.006^{**}$ \\
RF Cls & -2.069 & $0.033^{*}$ \\
RF Reg & -1.883 & $0.045^{*}$ \\
GB Cls & -1.123 & 0.144 \\
GB Reg & -1.098 & 0.149 \\
\hline
\end{tabular}
\end{minipage}

\vspace{1ex}

\begin{tabular}{@{}l@{}}
\footnotesize $\dag$ $p<.10$, $*$ $p<.05$, $**$ $p<.01$
\end{tabular}

\end{table*}

\begin{table*}[h]
\centering
\caption{HLN test results in the case of calculating $\beta$ from all results}\label{Tab4}

\begin{minipage}{0.45\textwidth}
\centering
\vspace{1ex}
\begin{tabular}{@{}lll@{}}
\hline
\multicolumn{3}{c}{uLSIF with three items} \\ 
\hline
Model & HLN stat & P-value \\
\hline
EN Cls  & -2.116 & $0.030^{*}$ \\
EN Reg  & -1.398 & $0.096^{\dag}$ \\
RF Cls  & -1.796 & $0.051^{\dag}$ \\
RF Reg  & -2.485 & $0.016^{*}$ \\
GB Cls  & -1.736 & $0.057^{\dag}$ \\
GB Reg  & -0.916 & 0.191 \\
\hline
\end{tabular}
\end{minipage}
\hfill
\begin{minipage}{0.52\textwidth}
\centering
\vspace{1ex}
\begin{tabular}{@{}lll@{}}
\hline
\multicolumn{3}{c}{uLSIF with all items} \\ 
\hline
Model & HLN stat & P-value \\
\hline
EN Cls  & -1.725 & $0.058^{\dag}$ \\
EN Reg  & -1.570 & $0.074^{\dag}$ \\
RF Cls  & -1.978 & $0.038^{*}$ \\
RF Reg  & -2.423 & $0.018^{*}$ \\
GB Cls  & -1.449 & $0.089^{\dag}$ \\
GB Reg  & -1.397 & $0.096^{\dag}$ \\
\hline
\end{tabular}
\end{minipage}

\vspace{1ex}

\begin{tabular}{@{}l@{}}
\footnotesize $\dag$ $p<.10$, $*$ $p<.05$, $**$ $p<.01$
\end{tabular}

\end{table*}

\section{Discussion and Conclusion}\label{sec7}
The objective of this study is to develop a framework for producing official statistics using new data sources that are subject to selection bias. In recent years, national statistical institutes have begun to incorporate non-traditional data sources—such as POS data and mobile phone GPS data—into the production of official statistics. These efforts are expected to enhance the quality of decision-making, including economic policymaking. Within this broader trend, it is a natural progression to utilize the vast amount of transaction data accumulated by private companies for official statistical purposes. However, the adoption of such data has been slower than anticipated.

This is primarily due to the significant selection bias inherent in these data when they are treated as sources for official statistics. That is, if such strong selection bias can be properly corrected, then large volumes of potentially valuable transaction data could be utilized as inputs for official statistics. This would substantially expand the scope of official statistical systems and improve the quality of a wide range of decision-making processes.

Using Japanese labor statistics as a case study, this research demonstrates that even data affected by selection bias can be used to construct useful preliminary indicators. As shown in Section \ref{sec51}, the indicator in question normally suffers from a publication lag of more than one year, making it fundamentally valuable but unusable for real-time decision-making. However, by applying the method proposed in this study, it becomes feasible to produce a timely version of this indicator that can support real-time decision-making.

As discussed in Section \ref{sec2}, this study replaces more basic existing research \cite{bib33} with a more realistic setting. The earlier study assumes that government agencies can obtain real-time data from private employment agencies and combine it with the latest official statistics to complete the estimation process within the government. In contrast, this study does not rely on this unrealistic assumption. At least in the context of Japan, it is challenging for government agencies to acquire real-time data from private employment agencies and complete the estimation internally. Realistically, a separate organization outside the government would need to integrate real-time data from private employment agencies with the latest available official statistics and perform the estimation. This would result in a significant time lag in the raw data from government statistics. This study demonstrated that reasonably accurate prompt release can be achieved under such conditions.

The contribution of this study is not limited to the labor statistics used in the experiment. The approach introduced in this research is adaptable to not just these labor statistics but also to a range of other data sets. There are thought to be many situations with the same structure.

To date, the application of machine learning techniques in traditional official statistical practices has been extremely limited. However, methodologies developed in the field of machine learning offer significant utility, particularly in the context of leveraging non-traditional data sources. If this research contributes to the wider dissemination of machine learning approaches in official statistics, it would represent a meaningful advancement in the field.

\backmatter

\bmhead{Acknowledgments}
We would like to express our appreciation to the Ministry of Health, Labour and Welfare and the Ministry of Internal Affairs and Communications for providing us with survey data as well as Recruit Holdings Co., Ltd. for providing us with transaction data of their service, Recruit Agent. Support provided by members of the laboratory to which we belong is gratefully acknowledged.

\bmhead{Data availability}
The transaction data from a private employment agency used in this study may be obtainable from Recruit Holdings Co., Ltd., but are not publicly available and require  individual negotiations as there is no established application process. The government survey samples used in this study may be obtainable from the Ministry of Health, Labor, and Welfare in Japan, but are not publicly available. Although there is an established application process for some non-open data, the data used in this study require individual negotiations. 

\section*{Declarations}

\bmhead{Conflict of interest}
The authors declare no conflict of interest.

\bmhead{Author contributions} Yuya Takada wrote the manuscript under the supervision of Kiyoshi Izumi.

\begin{itemize}
\item Funding: Not applicable
%\item Conflict of interest/Competing interests (check journal-specific guidelines for which heading to use)
\item Ethics approval: Not applicable
\item Consent to participate: Not applicable
\item Consent for publication: Not applicable
\item Availability of data and materials: Not applicable
\item Code availability: Not applicable
%\item Authors' contributions
\end{itemize}

\bibliographystyle{sn-mathphys}
\bibliography{sn-bibliography}


\begin{thebibliography}{29}
\ifx \bisbn   \undefined \def \bisbn  #1{ISBN #1}\fi
\ifx \binits  \undefined \def \binits#1{#1}\fi
\ifx \bauthor  \undefined \def \bauthor#1{#1}\fi
\ifx \batitle  \undefined \def \batitle#1{#1}\fi
\ifx \bjtitle  \undefined \def \bjtitle#1{#1}\fi
\ifx \bvolume  \undefined \def \bvolume#1{\textbf{#1}}\fi
\ifx \byear  \undefined \def \byear#1{#1}\fi
\ifx \bissue  \undefined \def \bissue#1{#1}\fi
\ifx \bfpage  \undefined \def \bfpage#1{#1}\fi
\ifx \blpage  \undefined \def \blpage #1{#1}\fi
\ifx \burl  \undefined \def \burl#1{\textsf{#1}}\fi
\ifx \doiurl  \undefined \def \doiurl#1{\url{https://doi.org/#1}}\fi
\ifx \betal  \undefined \def \betal{\textit{et al.}}\fi
\ifx \binstitute  \undefined \def \binstitute#1{#1}\fi
\ifx \binstitutionaled  \undefined \def \binstitutionaled#1{#1}\fi
\ifx \bctitle  \undefined \def \bctitle#1{#1}\fi
\ifx \beditor  \undefined \def \beditor#1{#1}\fi
\ifx \bpublisher  \undefined \def \bpublisher#1{#1}\fi
\ifx \bbtitle  \undefined \def \bbtitle#1{#1}\fi
\ifx \bedition  \undefined \def \bedition#1{#1}\fi
\ifx \bseriesno  \undefined \def \bseriesno#1{#1}\fi
\ifx \blocation  \undefined \def \blocation#1{#1}\fi
\ifx \bsertitle  \undefined \def \bsertitle#1{#1}\fi
\ifx \bsnm \undefined \def \bsnm#1{#1}\fi
\ifx \bsuffix \undefined \def \bsuffix#1{#1}\fi
\ifx \bparticle \undefined \def \bparticle#1{#1}\fi
\ifx \barticle \undefined \def \barticle#1{#1}\fi
\bibcommenthead
\ifx \bconfdate \undefined \def \bconfdate #1{#1}\fi
\ifx \botherref \undefined \def \botherref #1{#1}\fi
\ifx \url \undefined \def \url#1{\textsf{#1}}\fi
\ifx \bchapter \undefined \def \bchapter#1{#1}\fi
\ifx \bbook \undefined \def \bbook#1{#1}\fi
\ifx \bcomment \undefined \def \bcomment#1{#1}\fi
\ifx \oauthor \undefined \def \oauthor#1{#1}\fi
\ifx \citeauthoryear \undefined \def \citeauthoryear#1{#1}\fi
\ifx \endbibitem  \undefined \def \endbibitem {}\fi
\ifx \bconflocation  \undefined \def \bconflocation#1{#1}\fi
\ifx \arxivurl  \undefined \def \arxivurl#1{\textsf{#1}}\fi
\csname PreBibitemsHook\endcsname

%%% 1
\bibitem[\protect\citeauthoryear{{Task Teams: Mobile Phone Data}}{}]{bib8}
\begin{botherref}
\oauthor{\bsnm{{Task Teams: Mobile Phone Data}}}:
United Nations Statistics Division.
\url{https://unstats.un.org/bigdata/task-teams/mobile-phone/index.cshtml}
\end{botherref}
\endbibitem

%%% 2
\bibitem[\protect\citeauthoryear{Division}{2019}]{bib9}
\begin{bbook}
\bauthor{\bsnm{Division}, \binits{U.N.S.}}:
\bbtitle{Handbook on the Use of Mobile Phone Data for Official Statistics},
(\byear{2019})
\end{bbook}
\endbibitem

%%% 3
\bibitem[\protect\citeauthoryear{Kroon}{2012}]{bib11}
\begin{bchapter}
\bauthor{\bsnm{Kroon}, \binits{J.}}:
\bctitle{Mobile positioning as a possible data source for international travel
  service statistics}.
In: \bbtitle{UNECE Conference of European Statisticians}
(\byear{2012})
\end{bchapter}
\endbibitem

%%% 4
\bibitem[\protect\citeauthoryear{Feenstra and Shapiro}{2003}]{bib1}
\begin{bchapter}
\bauthor{\bsnm{Feenstra}, \binits{R.C.}},
\bauthor{\bsnm{Shapiro}, \binits{M.D.}}:
\bctitle{High-frequency substitution and the measurement of price indexes}.
In: \bbtitle{Scanner Data and Price Indexes},
pp. \bfpage{123}--\blpage{150}.
\bpublisher{University of Chicago Press}, \blocation{???}
(\byear{2003}).
\doiurl{10.3386/w8176}
\end{bchapter}
\endbibitem

%%% 5
\bibitem[\protect\citeauthoryear{Haan and der Grient}{2011}]{bib2}
\begin{barticle}
\bauthor{\bsnm{Haan}, \binits{J.D.}},
\bauthor{\bsnm{Grient}, \binits{H.A.V.}}:
\batitle{Eliminating chain drift in price indexes based on scanner data}.
\bjtitle{Journal of Econometrics}
\bvolume{161}(\bissue{1}),
\bfpage{36}--\blpage{46}
(\byear{2011})
\doiurl{10.1016/j.jeconom.2010.09.004}
\end{barticle}
\endbibitem

%%% 6
\bibitem[\protect\citeauthoryear{Ivancic et~al.}{2011}]{bib3}
\begin{barticle}
\bauthor{\bsnm{Ivancic}, \binits{L.}},
\bauthor{\bsnm{Diewert}, \binits{W.E.}},
\bauthor{\bsnm{Fox}, \binits{K.J.}}:
\batitle{Scanner data, time aggregation and the construction of price indexes}.
\bjtitle{Journal of Econometrics}
\bvolume{161}(\bissue{1}),
\bfpage{24}--\blpage{35}
(\byear{2011})
\doiurl{10.1016/j.jeconom.2010.09.003}
\end{barticle}
\endbibitem

%%% 7
\bibitem[\protect\citeauthoryear{Watanabe and Watanabe}{2014}]{bib4}
\begin{botherref}
\oauthor{\bsnm{Watanabe}, \binits{K.}},
\oauthor{\bsnm{Watanabe}, \binits{T.}}:
Estimating daily inflation using scanner data: A progress report.
Technical Report F-342,
CARF Working Paper
(2014)
\end{botherref}
\endbibitem

%%% 8
\bibitem[\protect\citeauthoryear{Cavallo}{2013}]{bib5}
\begin{barticle}
\bauthor{\bsnm{Cavallo}, \binits{A.}}:
\batitle{Online and official price indexes: Measuring argentina's inflation}.
\bjtitle{Journal of Monetary Economics}
\bvolume{60}(\bissue{2}),
\bfpage{152}--\blpage{165}
(\byear{2013})
\doiurl{10.1016/j.jmoneco.2012.10.002}
\end{barticle}
\endbibitem

%%% 9
\bibitem[\protect\citeauthoryear{Cavallo and Rigobon}{2016}]{bib6}
\begin{barticle}
\bauthor{\bsnm{Cavallo}, \binits{A.}},
\bauthor{\bsnm{Rigobon}, \binits{R.}}:
\batitle{The billion prices project: Using online prices for measurement and
  research}.
\bjtitle{Journal of Economic Perspectives}
\bvolume{30}(\bissue{2}),
\bfpage{151}--\blpage{78}
(\byear{2016})
\doiurl{10.1257/jep.30.2.151}
\end{barticle}
\endbibitem

%%% 10
\bibitem[\protect\citeauthoryear{Cavallo et~al.}{2018}]{bib7}
\begin{bchapter}
\bauthor{\bsnm{Cavallo}, \binits{A.}},
\bauthor{\bsnm{Diewert}, \binits{W.E.}},
\bauthor{\bsnm{Feenstra}, \binits{R.C.}},
\bauthor{\bsnm{Inklaar}, \binits{R.}},
\bauthor{\bsnm{Timmer}, \binits{M.P.}}:
\bctitle{Using online prices for measuring real consumption across countries}.
In: \bbtitle{AEA Papers and Proceedings},
vol. \bseriesno{108},
pp. \bfpage{483}--\blpage{87}
(\byear{2018}).
\doiurl{10.1257/pandp.20181003}
\end{bchapter}
\endbibitem

%%% 11
\bibitem[\protect\citeauthoryear{Woloszko}{2020}]{bib13}
\begin{botherref}
\oauthor{\bsnm{Woloszko}, \binits{N.}}:
Tracking activity in real time with google trends.
Technical report,
OECD Economics Department Working Papers
(2020).
\doiurl{10.1787/6b9c7518-en}
\end{botherref}
\endbibitem

%%% 12
\bibitem[\protect\citeauthoryear{Schiavoni et~al.}{2019}]{bib14}
\begin{botherref}
\oauthor{\bsnm{Schiavoni}, \binits{C.}},
\oauthor{\bsnm{Palm}, \binits{F.}},
\oauthor{\bsnm{Smeekes}, \binits{S.}},
\oauthor{\bsnm{Brakel}, \binits{J.V.D.}}:
A dynamic factor model approach to incorporate big data in state space models
  for official statistics.
Journal of the Royal Statistical Society Series A (Statistics in Society),
324--353
(2019)
\doiurl{10.1111/rssa.12417}
\end{botherref}
\endbibitem

%%% 13
\bibitem[\protect\citeauthoryear{Kaczmarek et~al.}{2025}]{bib38}
\begin{barticle}
\bauthor{\bsnm{Kaczmarek}, \binits{T.}},
\bauthor{\bsnm{Gajowniczek}, \binits{K.}},
\bauthor{\bsnm{Bat{\'o}g}, \binits{B.}}:
\batitle{Media sentiment and economic indicators: a case study of the polish
  economy during the covid-19 pandemic}.
\bjtitle{Journal of Computational Social Science}
(\byear{2025})
\doiurl{10.1007/s42001-025-00375-x}
\end{barticle}
\endbibitem

%%% 14
\bibitem[\protect\citeauthoryear{Wycoff et~al.}{2024}]{bib39}
\begin{barticle}
\bauthor{\bsnm{Wycoff}, \binits{S.}},
\bauthor{\bsnm{Alcorn}, \binits{N.}},
\bauthor{\bsnm{Pierson}, \binits{S.}},
\bauthor{\bsnm{Clinton}, \binits{W.}},
\bauthor{\bsnm{Smith}, \binits{E.}},
\bauthor{\bsnm{Elliott}, \binits{M.}}:
\batitle{Using publicly available organic data to forecast forced migration
  from ukraine}.
\bjtitle{Journal of Computational Social Science}
\bvolume{7},
\bfpage{527}--\blpage{547}
(\byear{2024})
\doiurl{10.1007/s42001-024-00304-4}
\end{barticle}
\endbibitem

%%% 15
\bibitem[\protect\citeauthoryear{Arcila-Calder{\'o}n et~al.}{2025}]{bib40}
\begin{barticle}
\bauthor{\bsnm{Arcila-Calder{\'o}n}, \binits{C.}},
\bauthor{\bsnm{Recuero}, \binits{R.}},
\bauthor{\bsnm{Said-Hung}, \binits{E.}},
\bauthor{\bsnm{Mar{\'i}n-Guti{\'e}rrez}, \binits{I.}},
\bauthor{\bsnm{Blanco-Herrero}, \binits{D.}},
\bauthor{\bsnm{Cano}, \binits{F.}}:
\batitle{Cross-border mobility and twitter data: Observing the turkish-european
  border crisis with mobile phones and social media}.
\bjtitle{Journal of Computational Social Science}
(\byear{2025})
\doiurl{10.1007/s42001-024-00354-8}
\end{barticle}
\endbibitem

%%% 16
\bibitem[\protect\citeauthoryear{Takada and Izumi}{2022}]{bib33}
\begin{bchapter}
\bauthor{\bsnm{Takada}, \binits{Y.}},
\bauthor{\bsnm{Izumi}, \binits{K.}}:
\bctitle{Implementation of biased big data to the japanese official labor
  statistics using supervised learning under covariate shift}.
In: \bbtitle{2022 IEEE International Conference on Big Data (Big Data)},
pp. \bfpage{2062}--\blpage{2071}
(\byear{2022}).
\doiurl{10.1109/BigData55660.2022.10020681}
\end{bchapter}
\endbibitem

%%% 17
\bibitem[\protect\citeauthoryear{Sugiyama et~al.}{2012}]{bib17}
\begin{bbook}
\bauthor{\bsnm{Sugiyama}, \binits{M.}},
\bauthor{\bsnm{Suzuki}, \binits{T.}},
\bauthor{\bsnm{Kanamori}, \binits{T.}}:
\bbtitle{Density Ratio Estimation in Machine Learning}.
\bpublisher{Cambridge University Press},
\blocation{Cambridge}
(\byear{2012}).
\doiurl{10.1017/CBO9781139035613}
\end{bbook}
\endbibitem

%%% 18
\bibitem[\protect\citeauthoryear{Shimodaira}{2000}]{bib28}
\begin{barticle}
\bauthor{\bsnm{Shimodaira}, \binits{H.}}:
\batitle{Improving predictive inference under covariate shift by weighting the
  log-likelihood function}.
\bjtitle{Journal of statistical planning and inference}
\bvolume{90}(\bissue{2}),
\bfpage{227}--\blpage{244}
(\byear{2000})
\doiurl{10.1016/S0378-3758(00)00115-4}
\end{barticle}
\endbibitem

%%% 19
\bibitem[\protect\citeauthoryear{Rubin}{1976}]{bib29}
\begin{barticle}
\bauthor{\bsnm{Rubin}, \binits{D.B.}}:
\batitle{Inference and missing data}.
\bjtitle{Biometrika}
\bvolume{63}(\bissue{3}),
\bfpage{581}--\blpage{592}
(\byear{1976})
\doiurl{10.1093/biomet/63.3.581}
\end{barticle}
\endbibitem

%%% 20
\bibitem[\protect\citeauthoryear{Yang et~al.}{2024}]{bib30}
\begin{botherref}
\oauthor{\bsnm{Yang}, \binits{Y.}},
\oauthor{\bsnm{Kuchibhotla}, \binits{A.K.}},
\oauthor{\bsnm{Tchetgen~Tchetgen}, \binits{E.}}:
Doubly robust calibration of prediction sets under covariate shift.
Journal of the Royal Statistical Society Series B: Statistical Methodology,
009
(2024)
\doiurl{10.1093/jrsssb/qkae009}
\end{botherref}
\endbibitem

%%% 21
\bibitem[\protect\citeauthoryear{Graham}{2009}]{bib31}
\begin{barticle}
\bauthor{\bsnm{Graham}, \binits{J.W.}}:
\batitle{Missing data analysis: Making it work in the real world}.
\bjtitle{Annual review of psychology}
\bvolume{60}(\bissue{1}),
\bfpage{549}--\blpage{576}
(\byear{2009})
\doiurl{10.1146/annurev.psych.58.110405.085530}
\end{barticle}
\endbibitem

%%% 22
\bibitem[\protect\citeauthoryear{Morikawa and Kim}{2021}]{bib32}
\begin{barticle}
\bauthor{\bsnm{Morikawa}, \binits{K.}},
\bauthor{\bsnm{Kim}, \binits{J.K.}}:
\batitle{Semiparametric optimal estimation with nonignorable nonresponse data}.
\bjtitle{The Annals of Statistics}
\bvolume{49}(\bissue{5}),
\bfpage{2991}--\blpage{3014}
(\byear{2021})
\doiurl{10.1214/20-AOS2053}
\end{barticle}
\endbibitem

%%% 23
\bibitem[\protect\citeauthoryear{H\"{a}rdle et~al.}{2004}]{bib34}
\begin{bbook}
\bauthor{\bsnm{H\"{a}rdle}, \binits{W.}},
\bauthor{\bsnm{M\"{u}ller}, \binits{M.}},
\bauthor{\bsnm{Sperlich}, \binits{S.}},
\bauthor{\bsnm{Werwatz}, \binits{A.}}, \betal:
\bbtitle{Nonparametric and Semiparametric Models}.
\bpublisher{Springer},
\blocation{Berlin}
(\byear{2004}).
\doiurl{10.1007/978-3-642-17147-0}
\end{bbook}
\endbibitem

%%% 24
\bibitem[\protect\citeauthoryear{Huang et~al.}{2006}]{bib35}
\begin{bchapter}
\bauthor{\bsnm{Huang}, \binits{J.}},
\bauthor{\bsnm{Gretton}, \binits{A.}},
\bauthor{\bsnm{Borgwardt}, \binits{K.}},
\bauthor{\bsnm{Sch\"{o}lkopf}, \binits{B.}},
\bauthor{\bsnm{Smola}, \binits{A.}}:
\bctitle{Correcting sample selection bias by unlabeled data}.
In: \bbtitle{Proceedings of the 19th International Conference on Neural
  Information Processing Systems},
pp. \bfpage{601}--\blpage{608}
(\byear{2006}).
\doiurl{10.5555/2976456.2976531}
\end{bchapter}
\endbibitem

%%% 25
\bibitem[\protect\citeauthoryear{Qin}{1998}]{bib21}
\begin{barticle}
\bauthor{\bsnm{Qin}, \binits{J.}}:
\batitle{Inferences for case-control and semiparametric two-sample density
  ratio models}.
\bjtitle{Biometrika}
\bvolume{85}(\bissue{3}),
\bfpage{619}--\blpage{630}
(\byear{1998})
\doiurl{10.1093/biomet/85.3.619}
\end{barticle}
\endbibitem

%%% 26
\bibitem[\protect\citeauthoryear{Cheng and Chu}{2004}]{bib22}
\begin{barticle}
\bauthor{\bsnm{Cheng}, \binits{K.F.}},
\bauthor{\bsnm{Chu}, \binits{C.K.}}:
\batitle{Semiparametric density estimation under a two-sample density ratio
  model}.
\bjtitle{Bernoulli}
\bvolume{10}(\bissue{4}),
\bfpage{583}--\blpage{604}
(\byear{2004})
\doiurl{10.3150/bj/1093265631}
\end{barticle}
\endbibitem

%%% 27
\bibitem[\protect\citeauthoryear{Bickel et~al.}{2007}]{bib36}
\begin{bchapter}
\bauthor{\bsnm{Bickel}, \binits{S.}},
\bauthor{\bsnm{Br\"{u}ckner}, \binits{M.}},
\bauthor{\bsnm{Scheffer}, \binits{T.}}:
\bctitle{Discriminative learning for differing training and test
  distributions}.
In: \bbtitle{Proceedings of the 24th International Conference on Machine
  Learning},
pp. \bfpage{81}--\blpage{88}
(\byear{2007}).
\doiurl{10.1145/1273496.1273507}
\end{bchapter}
\endbibitem

%%% 28
\bibitem[\protect\citeauthoryear{Sugiyama et~al.}{2008}]{bib37}
\begin{barticle}
\bauthor{\bsnm{Sugiyama}, \binits{M.}},
\bauthor{\bsnm{Suzuki}, \binits{T.}},
\bauthor{\bsnm{Nakajima}, \binits{S.}},
\bauthor{\bsnm{Kashima}, \binits{H.}},
\bauthor{\bsnm{B\"{u}nau}, \binits{P.}},
\bauthor{\bsnm{Kawanabe}, \binits{M.}}:
\batitle{Direct importance estimation for covariate shift adaptation}.
\bjtitle{Annals of the Institute of Statistical Mathematics}
\bvolume{60}(\bissue{4}),
\bfpage{699}--\blpage{746}
(\byear{2008})
\doiurl{10.1007/s10463-008-0197-x}
\end{barticle}
\endbibitem

%%% 29
\bibitem[\protect\citeauthoryear{Kanamori et~al.}{2009}]{bib25}
\begin{barticle}
\bauthor{\bsnm{Kanamori}, \binits{T.}},
\bauthor{\bsnm{Hido}, \binits{S.}},
\bauthor{\bsnm{Sugiyama}, \binits{M.}}:
\batitle{A least-squares approach to direct importance estimation}.
\bjtitle{The Journal of Machine Learning Research}
\bvolume{10},
\bfpage{1391}--\blpage{1445}
(\byear{2009})
\doiurl{10.5555/1577069.1755848}
\end{barticle}
\endbibitem

\end{thebibliography}

\end{document}